\documentclass[aps,prb,twocolumn,groupedaddress,showpacs]{revtex4}
\usepackage{color}
\usepackage{booktabs}
\usepackage{graphicx}
\usepackage{epsfig}
\usepackage{amssymb}
\usepackage{bm}
\usepackage{graphpap}
\usepackage[normalem]{ulem}
\usepackage{hyperref}
\usepackage[all]{hypcap}
\newcommand{\cblu}{\color{blue}}

\hypersetup{colorlinks=true,citecolor=blue,linkcolor=red}

\begin{document}
\newdimen\heavyrulewidth
\newdimen\lightrulewidth
\newdimen\cmidrulewidth
\newdimen\belowrulesep
\newdimen\belowbottomsep
\newdimen\aboverulesep
\newdimen\abovetopsep
\newdimen\defaultaddspace
\heavyrulewidth=.08em
\lightrulewidth=.05em
\belowrulesep=.65ex
\belowbottomsep=0pt
\aboverulesep=.4ex
\abovetopsep=0pt
\defaultaddspace=.5em

\title{Silicon Nanocrystallites in SiO$_2$ Matrix: The Role of Disorder and Size}

\author{Roberto \surname{Guerra}, Ivan \surname{Marri}, Rita
\surname{Magri}, Layla \surname{Martin-Samos}}
\affiliation{CNR-INFM-$S{^3}$ and Dipartimento di Fisica -
Universit\`a di Modena e Reggio Emilia - via Campi 213/A I-41100
Modena Italy.}
\author{Olivia  \surname{Pulci}}
\affiliation{European Theoretical Spectroscopy Facility (ETSF) and
CNR-INFM, Dept. of Physics, Universit\'a di Roma "Tor Vergata" Via
della Ricerca Scientifica 1, I-00133 Roma, Italy}
\author{Elena \surname{Degoli} and Stefano \surname{Ossicini}}
\affiliation{CNR-INFM-$S{^3}$ and Dipartimento di Scienze e Metodi
dell'Ingegneria - Universit\`a di Modena e Reggio Emilia - via
Amendola 2 Pad. Morselli, I-42100 Reggio Emilia Italy.}
\date{\today}

\begin{abstract}
We compare, through first-principles pseudopotential calculations,
the structural, electronic and optical properties of different size
silicon nanoclusters embedded in a SiO$_2$ crystalline or amorphous
matrix, with that of free-standing, hydrogenated and hydroxided
silicon nanoclusters of corresponding size and shape. We find that
the largest effect on the opto-electronic behavior is due to the
amorphization of the embedded nanocluster. In that, the
amorphization reduces the fundamental gap while increasing the
absorption strength in the visible range. Increasing the nanocluster
size does not change substantially this picture but only
leads to the reduction of the absorption threshold, following the quantum
confinement rule. Finally, through the calculation of the optical
absorption spectra both in a indipendent-particle and many-body
approach, we show that the effect of local fields is crucial for
describing properly the optical behavior of the crystalline case
while it is of minor importance for amorphous systems.
\end{abstract}

\maketitle

\section{Introduction}
It is generally accepted that the quantum confinement (QC),
observed when electron and hole are constrained in one or more
dimensions by a potential well, is essential for the visible light
emission in silicon nanoclusters (NCs) \cite{pavesi,ossospringer,bisi}.
Nevertheless, some open questions still exist.
For instance, it is not yet known how much {the} QC model correctly
describes the dependence of the energy gap $E_G$ on the NC size
and how {structure} distortions and surface properties can
influence the band-gap value.
Again, controversial interpretations of the photoluminescence (PL)
microscopic origin in low-dimensional Si structures still exist.
Surface effects, in particular oxidation\cite{ramos,ramos2,pavel},
on the NC optical properties have been addressed in the last years.
Many theoretical\cite{Wolkin,lup1,zhou} and experimental\cite{Wolkin}
works have proved that interface properties have dramatic effects on
the emission spectra of NCs. Besides, the size of the NC is a crucial
parameter\cite{kelires,kelires1} to determine whenever such
interface effects dominate (smaller NCs) or become negligible (larger NCs) respect to QC.\\
Various efforts have been done in order to describe optical
absorption and emission spectra in such systems. In this context,
inclusion of excitonic and many-body effects appears {to be} of fundamental importance\cite{louie}.
However, until now, {no work has yet tackled} the excitonic
problem for this kind of systems due to the difficulty to deal with a large number of atoms.

In 2003 the first theoretical {\it ab-initio} calculation considering
a small Si-NC embedded in a SiO$_2$ matrix \cite{daldosso,lup2} has
been published. In this work, both the host matrix and the embedded
nanostructure were fully relaxed.\\
In the present work a more detailed first-principles calculation
of the structural, electronic and optical properties of Si-NC
embedded in a SiO$_2$ matrix is reported. The two complementary
cases of a perfectly crystalline and a completely amorphous system
are taken into account. In particular, the effects generated by
the surrounding matrix on the electronic properties of {the} NCs
(of different size) are analyzed. The results are here compared
with the corresponding ones of the free standing systems of same
size.

The paper is organized as follows. A description of the theoretical
methods and of the systems is given in section \ref{sec_method}.
The structural, electronic and optical properties are analyzed
in section \ref{sec_GS} for both crystalline (section \ref{subsec_crystal})
and amorphous (section \ref{subsec_amorph}) cases. The effects of the
NCs size are analyzed in sec. \ref{subsec_size}.
{The} excited state properties described through many-body methods are reported
in section \ref{sec_EXC}. Conclusions are presented in section \ref{sec_concl}.

\section{The Method}\label{sec_method}
The $\beta$-cristobalite (BC) SiO$_2$ is well known to give rise to one of the
simplest Si/SiO$_2$ interface because of its diamond-like structure \cite{BC}.
The crystalline structure has been obtained from a Si$_{64}$O$_{128}$ cubic matrix
(of size L$=14.32$ \AA) by removing all oxygens included in a sphere of radius
4.4 \AA ~placed at the center of the cubic supercell.
The results is a structure of 64 Si and 116 O atoms with 10 Si bonded together
to form a small crystalline skeleton with T$_d$ local symmetry before relaxation.
In such core, Si atoms show a larger  bond length (3.1 \AA) with respect to that of the  Si bulk
structure (2.35 \AA). No defects (dangling bonds) are present, and all the O atoms
at the Si/SiO$_2$ interface are single bonded with the Si atoms of the NC.\\
\begin{figure}[b]\begin{center}
\centerline{\psfig{file=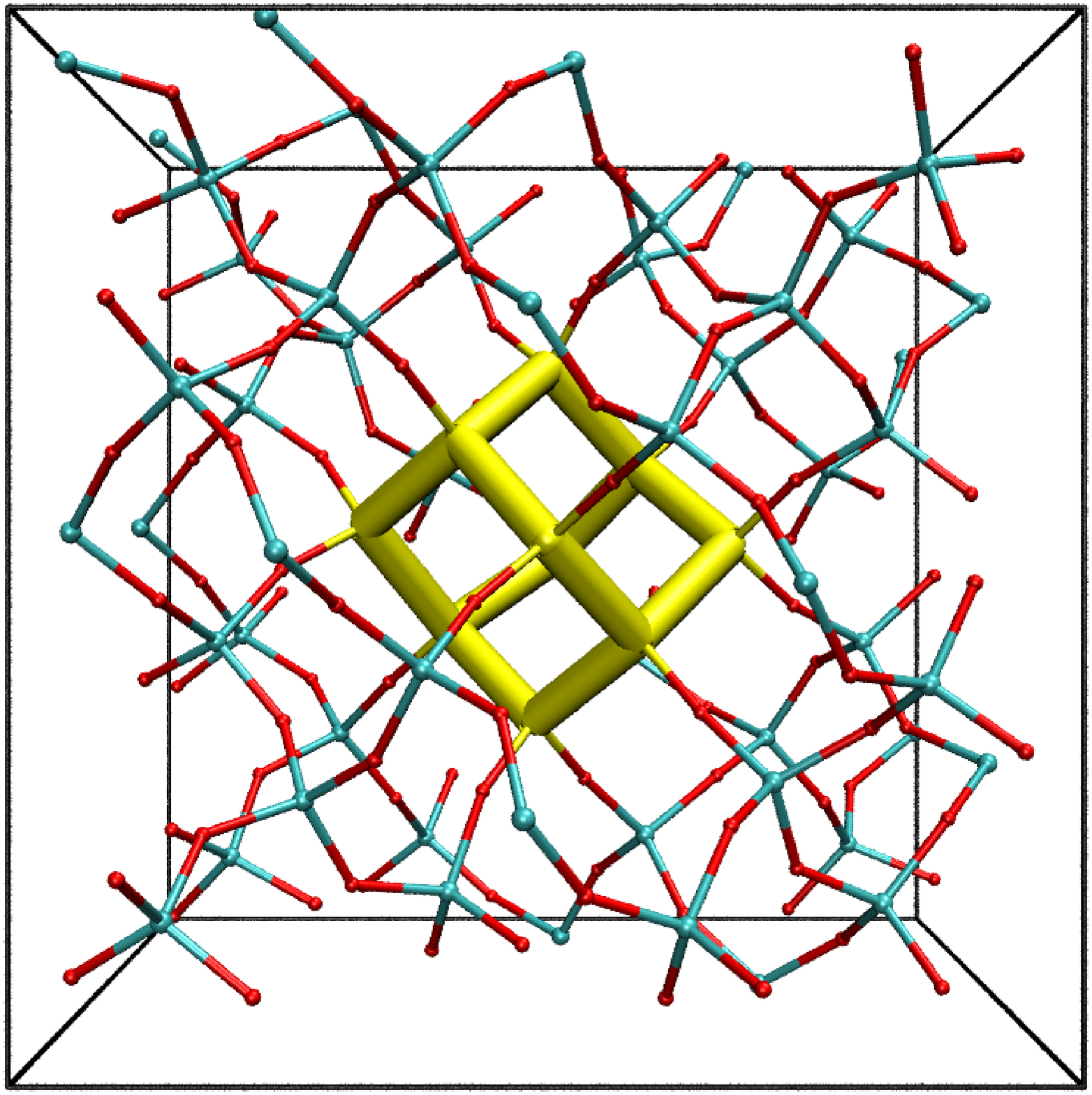,width=4.5cm}\psfig{file=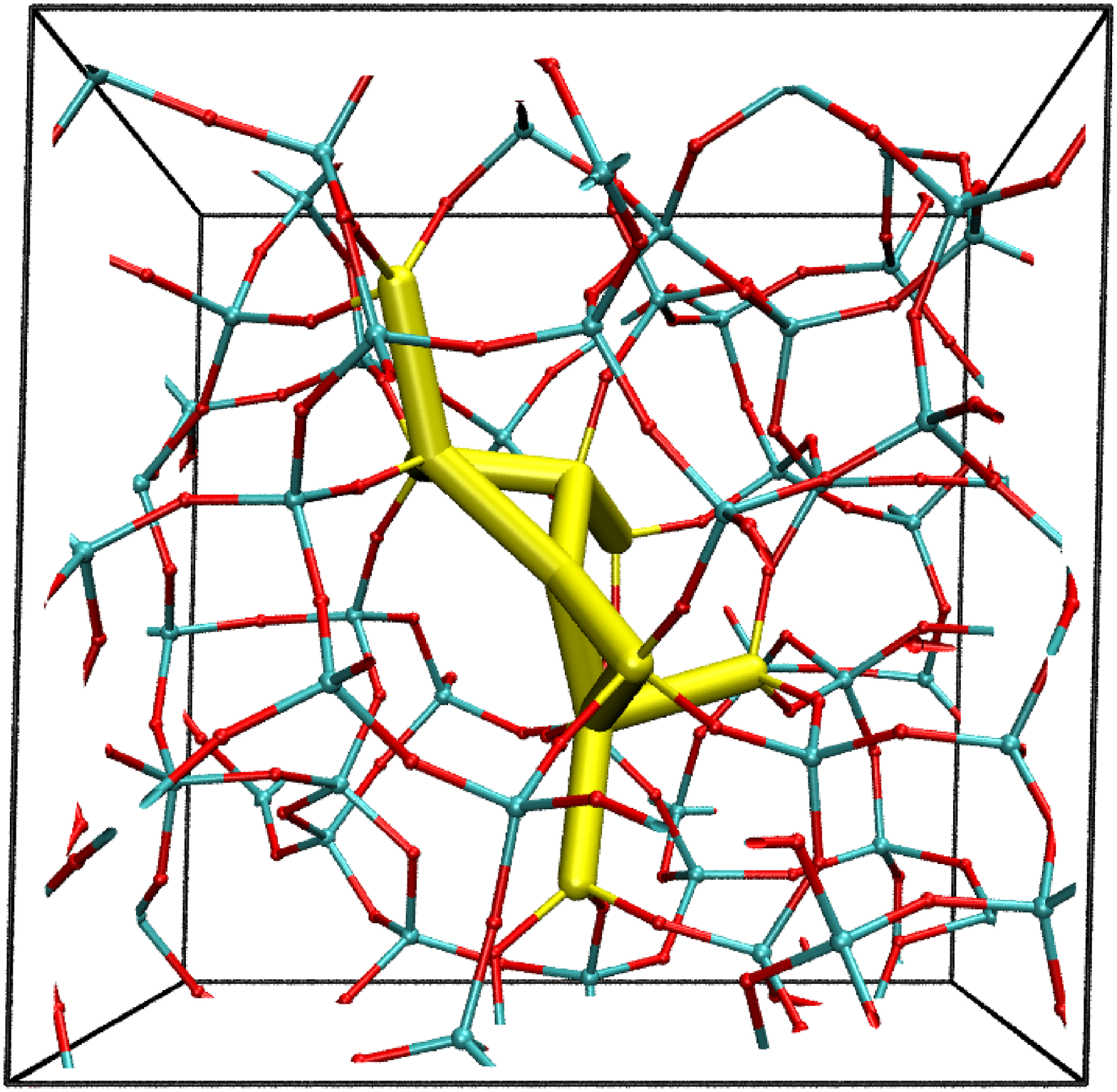,width=4.5cm}}
\caption{(color online) Stick and ball pictures of the final optimized structure of
    Si$_{10}$ in $\beta$-cristobalite matrix (left panel) or in a-SiO$_2$
    glass (right panel). Red (dark gray) spheres represent the O
    atoms, cyan (gray) spheres represent the Si of the matrix,
    and the yellow (gray) thick sticks represent the Si atoms of the NC.}
\label{struttura}\end{center}\end{figure}
The optimized structure has been achieved by relaxing the total volume
of the cell (see Fig.\ref{struttura}, left panel).
This approach gives a correct description of the atomic average density in the NC region.
Moreover, the small distortion induced in the host matrix
(essentially due to the metastable nature of the BC) is
practically irrelevant on the NC and the interface configuration.
On the other hand, relaxation criteria based on a fixed
supercell volume produce, when oxygens are removed, a strong
reduction of the atomic average density, giving thus not
realistic results. Together with the crystalline structure, the
complementary case of an amorphous silica (a-SiO$_2$) has been
considered. The glass model has been generated using classical
molecular dynamics (MD) simulations of quenching from a melt, as
described in Ref. \cite{layla}. The simulations have been done
using semi-empirical ionic potentials \cite{Beest}, assuming an
effective quench rate of $2.6 \cdot 10^{13}$K/s. The glass match
the well-connected definition of Zachariasen \cite{Zachariasen}.
The amorphous dot structure has been obtained starting from a
Si$_{64}$O$_{128}$ amorphous silica cell by removing the 10 oxygen
atoms included in a sphere of radius 3.5 {\AA}~ placed {at} the
center of the cell, as shown in Fig. \ref{struttura} (right
panel). The relaxation of all the structures have been performed
using the SIESTA code\cite{siesta1,siesta2} and Troullier-Martins
pseudopotentials with non-linear core corrections. A cutoff of
$38 Ry$ on the plane-wave basis set and no additional external pressure
or stress were applied. Atomic positions and cell parameters have
been left totally free to move.\\
\begin{figure}[t]\begin{center}
\centerline{\psfig{figure=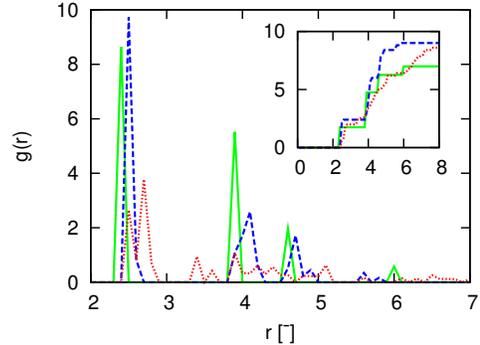,width=4.8cm,angle=270}}
\caption{(color online) Radial distribution function (and in the inset its integral) for the
    bulk silicon (green solid line), the crystalline Si$_{10}$ (blue dashed line)
    and amorphous a-Si$_{10}$ (red dotted line) NCs without the matrix contribution.}
\label{gr}\end{center}\end{figure} Figure \ref{gr} shows the
radial distribution function g(r) of the crystalline (Si$_{10}$)
and amorphous (a-Si$_{10}$) silicon NCs without considering
the matrix contribution, compared with that of the NC made
of the 8 atoms forming the primitive cubic cell of bulk silicon.
In the inset the integrated radial distribution functions are shown.
We note that the crystalline and the amorphous NCs present a
nearest-neighbor distance of about 2.43 \AA, strained respect to the
typical Si-bulk value (2.35 \AA), in fair agreement with the outcomes
of Yilmaz et al. \cite{yilmaz}. As expected, there is a good matching
between the g(r) of the crystalline NC and the Si-bulk
case, while the long-range order is clearly broken in the amorphous system.\\
Electronic and optical properties of the relaxed structures
have been obtained in the framework of DFT, using the ESPRESSO package\cite{espresso}.
Calculations have been performed using norm-conserving pseudopotentials within
the LDA approximation with a Ceperley-Alder exchange-correlation potential,
as parametrized by Perdew-Zunger. An energy cutoff of 60 Ry on the plane wave basis have been considered.
To include the many-body corrections we have {first} calculated, starting from the
DFT-LDA eigenvalues and eigenvectors, the GW\cite{G0W0} quasiparticle band
structure using 10975 plane waves for the correlation part of the self-energy,
and 20001 plane waves for the exchange part. The Bethe-Salpeter equation has been
{then} solved considering an excitonic Hamiltonian made up of more than
65.000 transitions. The Haydock algorithm has been used to solve the Hamiltonian.
Calculations have been performed using the EXC code\cite{exc_code}.

\section{Indipendent-Particle Results} \label{sec_GS}
In parallel to the crystalline and amorphous Si$_{10}$/SiO$_2$ systems,
three other structures have been studied:
(i) the pure {SiO$_2$} matrix (in the same phases
{as} used for the embedded NC calculations),
(ii) the isolated NCs as extracted from the relaxed NC-silica complex{es}
and capped by hydrogen atoms (Si$_{10}$-H),
(iii) the NCs together with the first interface oxygens extracted as in point (ii),
and then passivated by hydrogen atoms (Si$_{10}$-OH).
In the last two cases only {the} hydrogen atoms have been relaxed.
The goal is to distinguish between {the} properties that depend only
on the embedded NC from those that are instead
influenced by the presence of the matrix.
The comparison of the results relative to different
passivation regimes (H or OH groups) could give some insight
on the role played by the interface region.\\

\subsection{The crystalline cluster in $\beta$-cristobalite matrix} \label{subsec_crystal}

\begin{figure}[b]\begin{center}
\centerline{
	\psfig{figure=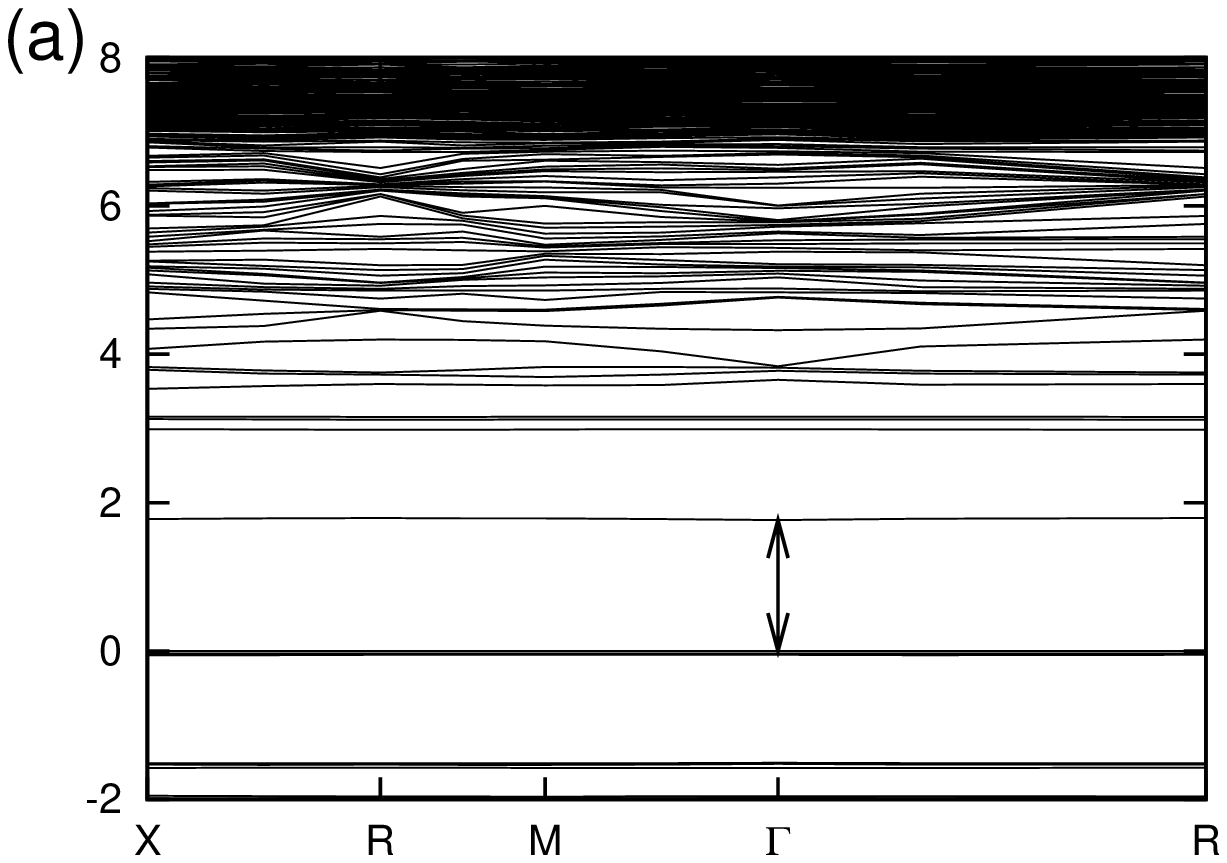,width=3.2cm,angle=270}
        \psfig{figure=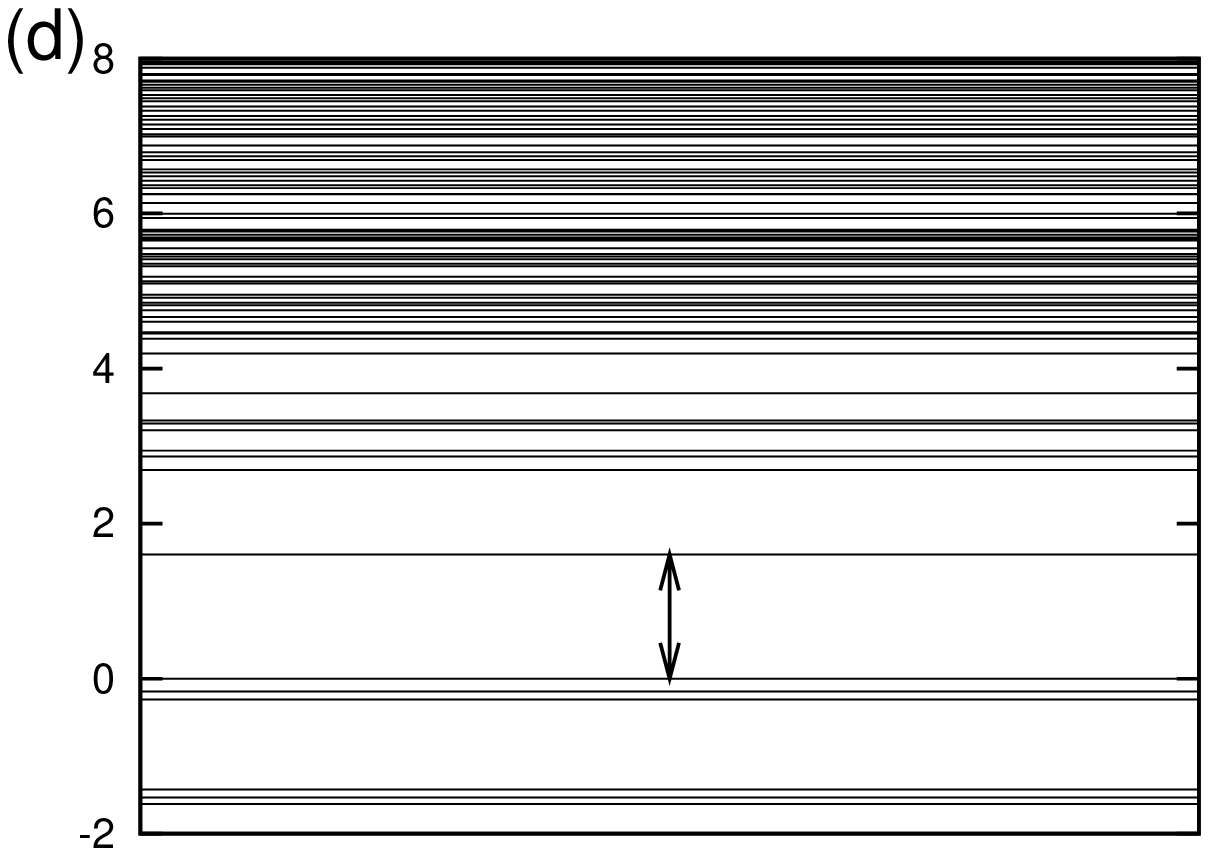,width=3.2cm,angle=270}}
\centerline{
	\psfig{figure=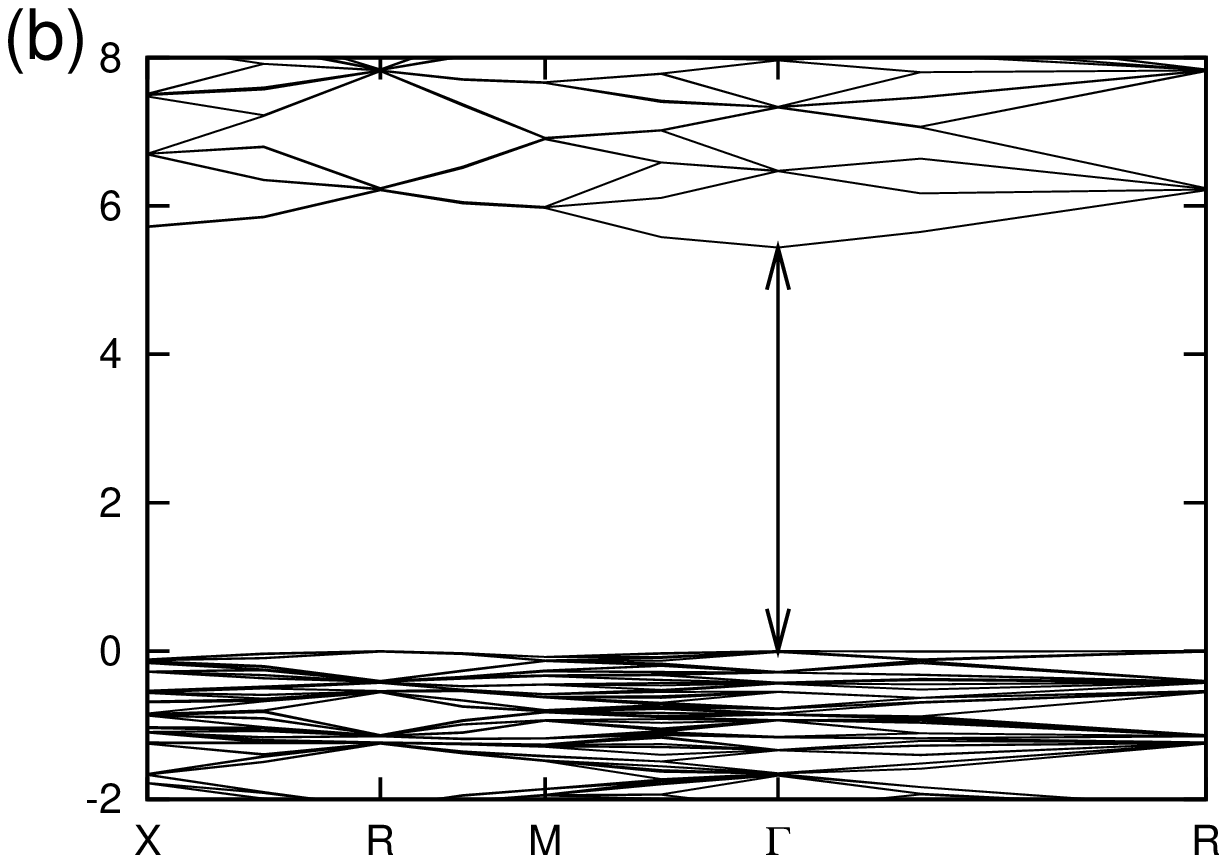,width=3.2cm,angle=270}
	\psfig{figure=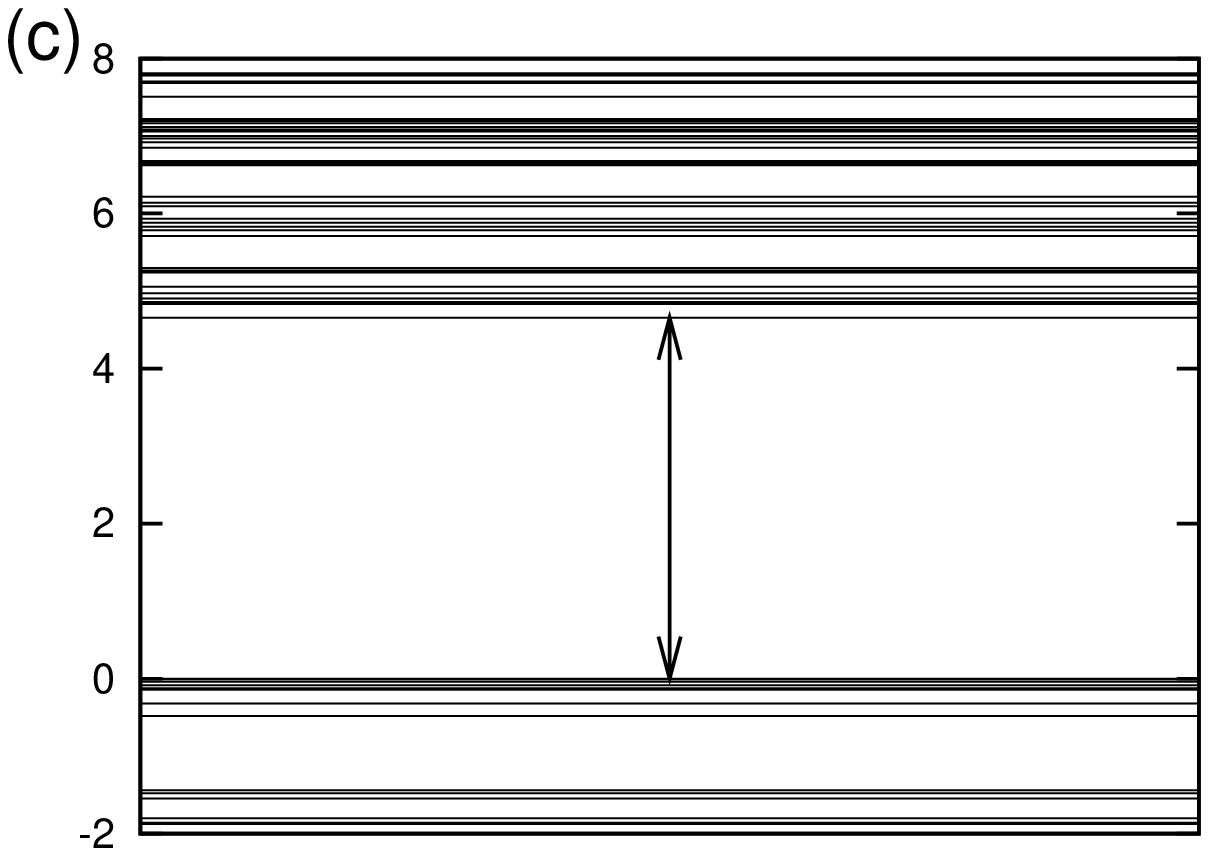,width=3.2cm,angle=270}}
\caption{Band structure along high symmetry points of the BZ for the crystalline Si$_{10}$
    NC in $\beta$-cristobalite (a), compared with the band structure of
    $\beta$-cristobalite bulk (b), the energy levels for the isolated Si$_{10}$
    NC passivated by H (c) and by OH groups (d).
    The arrows show the HOMO-LUMO band gap values. The units are in eV.}
\label{BC_bande}\end{center} \end{figure}
\begin{figure}[b]\begin{center}
\psfig{file=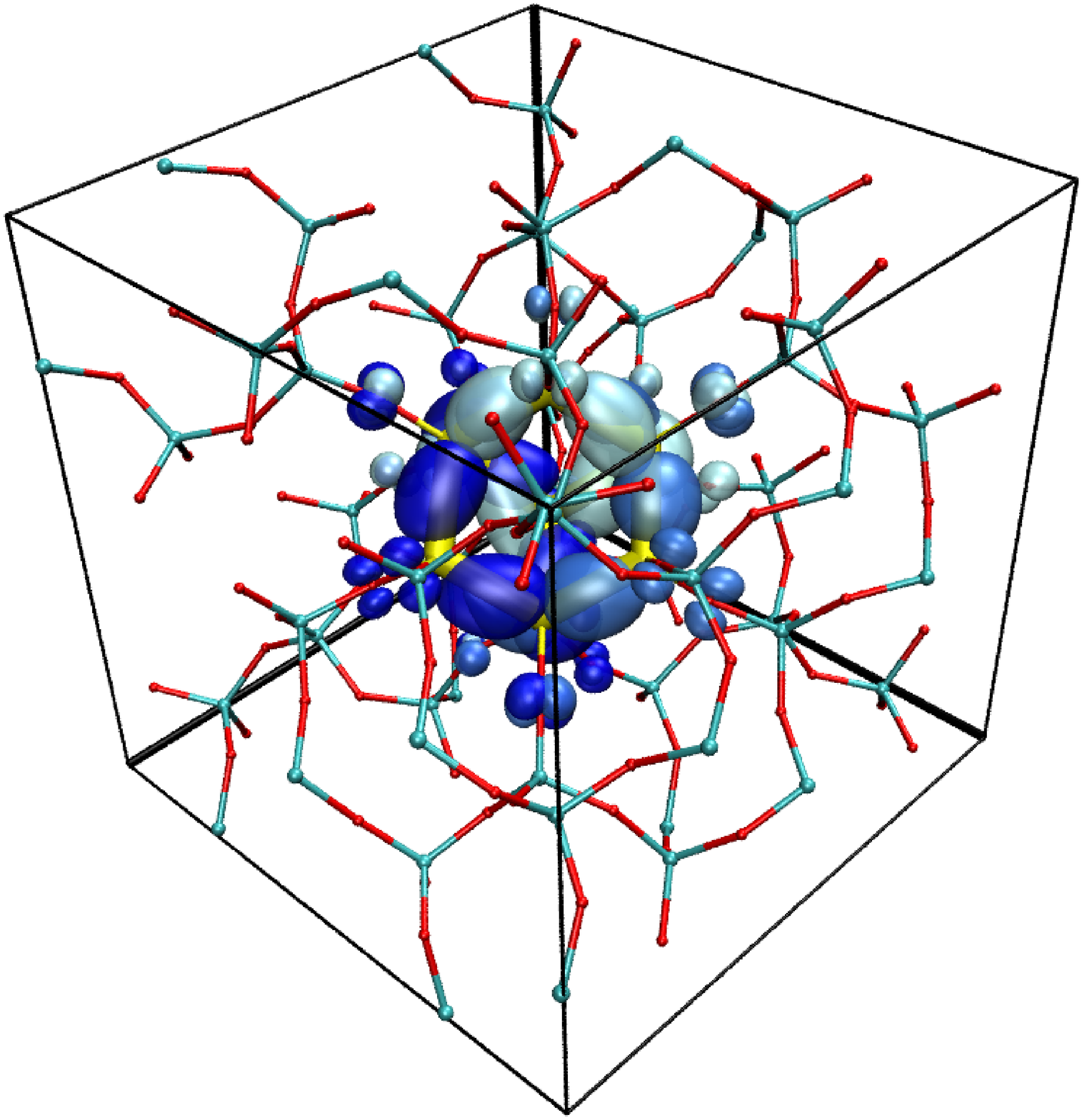,width=8cm}\\
\psfig{file=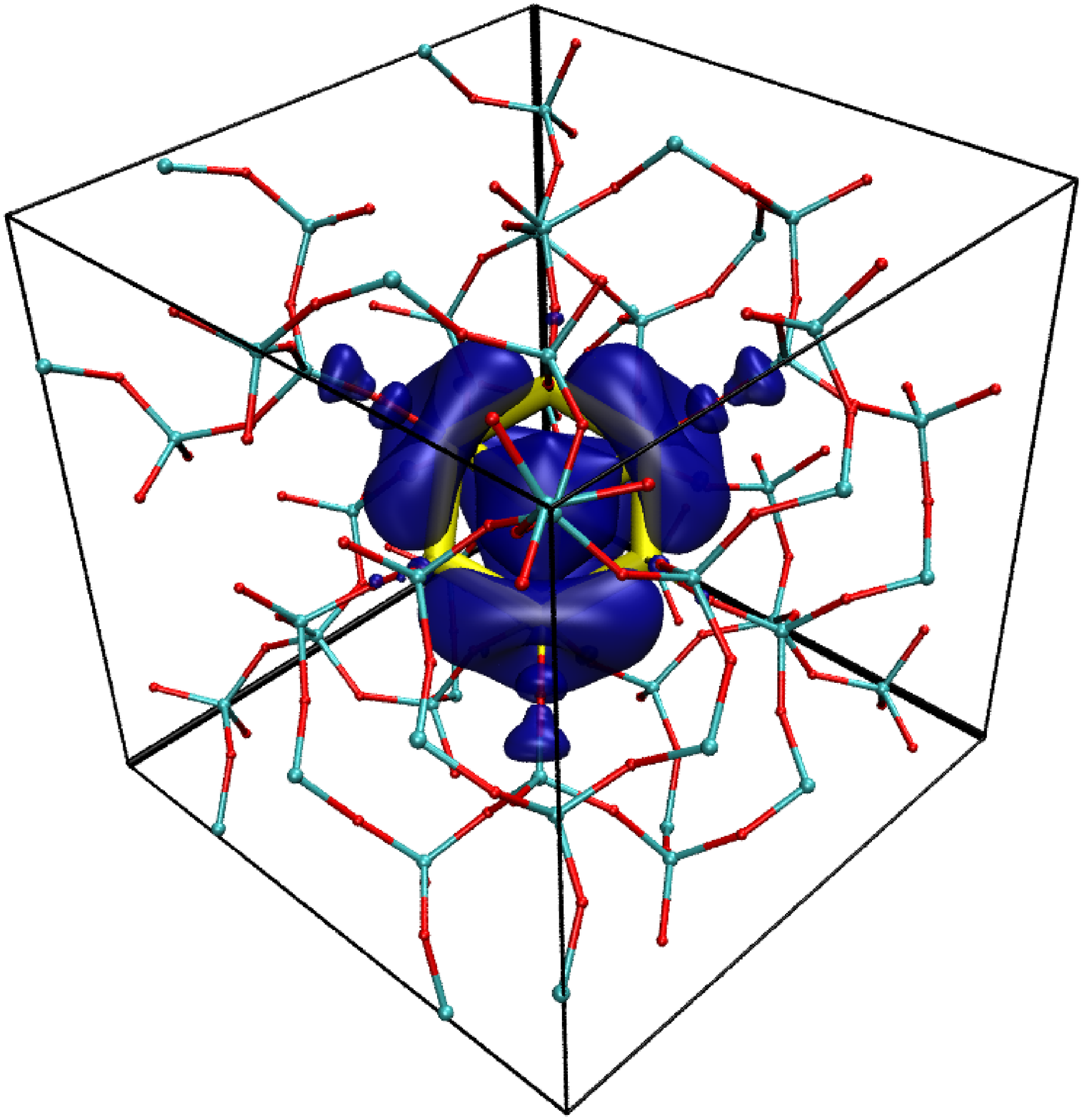,width=8cm}
\caption{(color online) Kohn-Sham orbitals at 10$\%$ of their max. amplitude
    for the Si$_{10}$ crystalline embedded NC.
    Top panel: HOMO state is represented in blue (gray), HOMO-1 in azure (light gray),
    HOMO-2 in lightest blue (lightest gray). Bottom panel: LUMO is represented in dark blue (dark gray).}
\label{HLBC}\end{center}\end{figure}

From the analysis of the relaxed Si$_{10}$/SiO$_2$ supercell
emerges that the NC has a strained structure with respect to bulk
Si \cite{yilmaz,Kroll,Bulutay} (see Fig. \ref{gr}), while the
$\beta$-cristobalite matrix is strongly distorted near the NC and
reduces progressively its stress far away from the
interface\cite{Watanabe2}. The electronic and optical properties
of the relaxed structures of Si$_{10}$/SiO$_2$, SiO$_2$ bulk,
Si$_{10}$-H and Si$_{10}$-OH have been computed. Figure
\ref{BC_bande} shows the comparison between the band structures
(energy levels) of all these systems. We see, first of all, a
strong reduction of the Si$_{10}$/SiO$_2$ energy gap ($E_G$=1.77
eV) with respect to both bulk SiO$_2$ (5.44 eV) and {the} isolated
NC passivated by H atoms (4.66 eV); finally, as confirmed by
Refs.~{\cblu \cite{ramos,ramos2,Wolkin}}, the passivation by OH
groups tends to red-shift the energy spectrum ($E_G$=1.6 eV). The
smaller gap with respect to the SiO$_2$ bulk case is clearly due
to the formation of confined states within the NC, lying between
the SiO$_2$ band edges, as displayed in Fig. \ref{BC_bande}a and
\ref{BC_bande}b. These states are fundamentally different from
those of the free-standing hydrogenated NC (Fig. \ref{BC_bande}c).
In the embedded NC states, indeed, we find a strong contribution
due to the presence of interfacial oxygen atoms. This is also
confirmed by the resemblance existing between the gap values of
the embedded NC and of the NC simply passivated by the OH groups
(Fig. \ref{BC_bande}d). $E_G$ is thus almost completely determined
by the barrier provided by the first shell of oxygen atoms. The
typical behavior of the bulk dispersed states is still recognizable
far away from the band edges, where the NC influence tends to vanish.
The square modulus contour plots of the highest occupied molecular orbital (HOMO)
and the two states just below it (HOMO-1, HOMO-2), for the Si$_{10}$
crystalline NC in the BC matrix are reported in  Fig. \ref{HLBC}.
The supercell relaxation breaks the symmetry of the system, and consequently
the degeneracy of the three states at the top of the valence band.
Therefore these states present slightly different eigenvalues but
preserve the character of the original symmetry.
In accordance with Ref.~{\cblu \cite{Kroll}}, these states are mainly
localized on part of the NC, in particular on the Si-Si bonds and
on the interface oxygens, while the lowest unoccupied molecular orbital
(LUMO) is localized on the interface and over all the entire NC region.
It is clearly shown {in Fig. \ref{HLBC}} that the first shell of oxygens around the NC are able to
completely trap the NC density charge, thus forming a strong barrier that is responsible
for the QC effect.

\begin{figure}[t]\begin{center}
\centerline{\psfig{figure=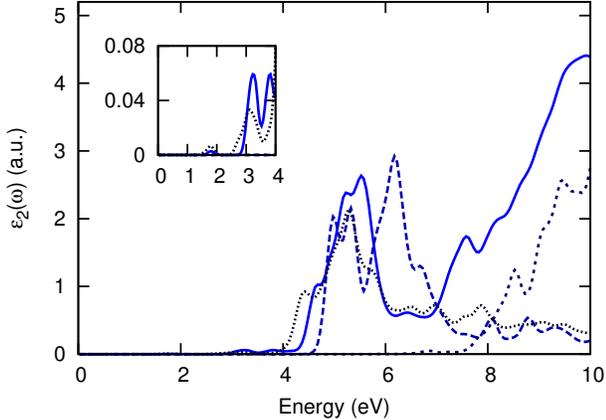,width=6cm,angle=270}}
\caption{(color online) DFT-RPA imaginary part of the dielectric function for the Si$_{10}$ crystalline NC
    in $\beta$-cristobalite matrix (solid line) compared with that of isolated
    Si$_{10}$-H NC (long-dashed line), the beta-cristobalite bulk (short-dashed line),
    and the Si$_{10}$-OH NC (dotted line). The inset shows an enlargement of the spectra at low energies.}
\label{eps2_crystal}\end{center}\end{figure}

In Figure \ref{eps2_crystal} the imaginary part of the dielectric
functions (calculated within the DFT-RPA approach, using the dipole approximation and
neglecting the local-field (LF) effects) {of} Si$_{10}$/SiO$_2$, Si$_{10}$-H and Si$_{10}$-OH NCs,
and {of} the BC bulk  are reported.
A detailed analysis of the results allows {to identify}  three distinct zones in the spectra.
The first one, above 6 eV, where the Si$_{10}$/SiO$_2$ and BC bulk are very similar.
The second one, between $\sim$4 eV and $\sim$6 eV, where a clear contribution due to
the embedded Si-NC is present. {The third one},
below 4 eV, where {the} transitions can not be ascribed neither to the matrix region nor to the pure NC.
In this {region the} peaks are originated by the interplay between the embedded NC
and the matrix; in particular they are due to the presence of
{the} oxygen atoms located at the interface region. This is
confirmed by the {similarity} that exists, in this energy region,
between the Si$_{10}$/SiO$_2$ and the Si$_{10}$-OH spectra, as
depicted in the insertion of Fig. \ref{eps2_crystal}. The presence
of OH groups thus induces a complex structure in the part of the
spectrum below 4 eV that can be compared with the peaks of the
Si$_{10}$/SiO$_2$ structures.

\subsection{The amorphous cluster in a glass} \label{subsec_amorph}

\begin{table}[t]\begin{center}\begin{tabular}[t]{c @{\hspace{0.6cm}} c @{\hspace{0.3cm}} c @{\hspace{0.3cm}} c @{\hspace{0.3cm}} c}
\toprule
            & SiO$_2$ & Si$_{10}$/SiO$_2$ & Si$_{10}$-OH & Si$_{10}$-H \\
\midrule
    Crystalline & 5.44    & 1.77              & 1.60         & 4.66 \\
    Amorphous   & 5.40    & 1.41              & 1.55         & 1.87 \\
\bottomrule
\end{tabular}
\caption{DFT HOMO-LUMO gap values (in eV) for the crystalline and amorphous
    silica, embedded, OH-terminated, H-terminated Si$_{10}$ NC.}
\label{tabella_gap1}\end{center}\end{table}
\begin{figure}[b]\begin{center}
\centerline{
	\psfig{figure=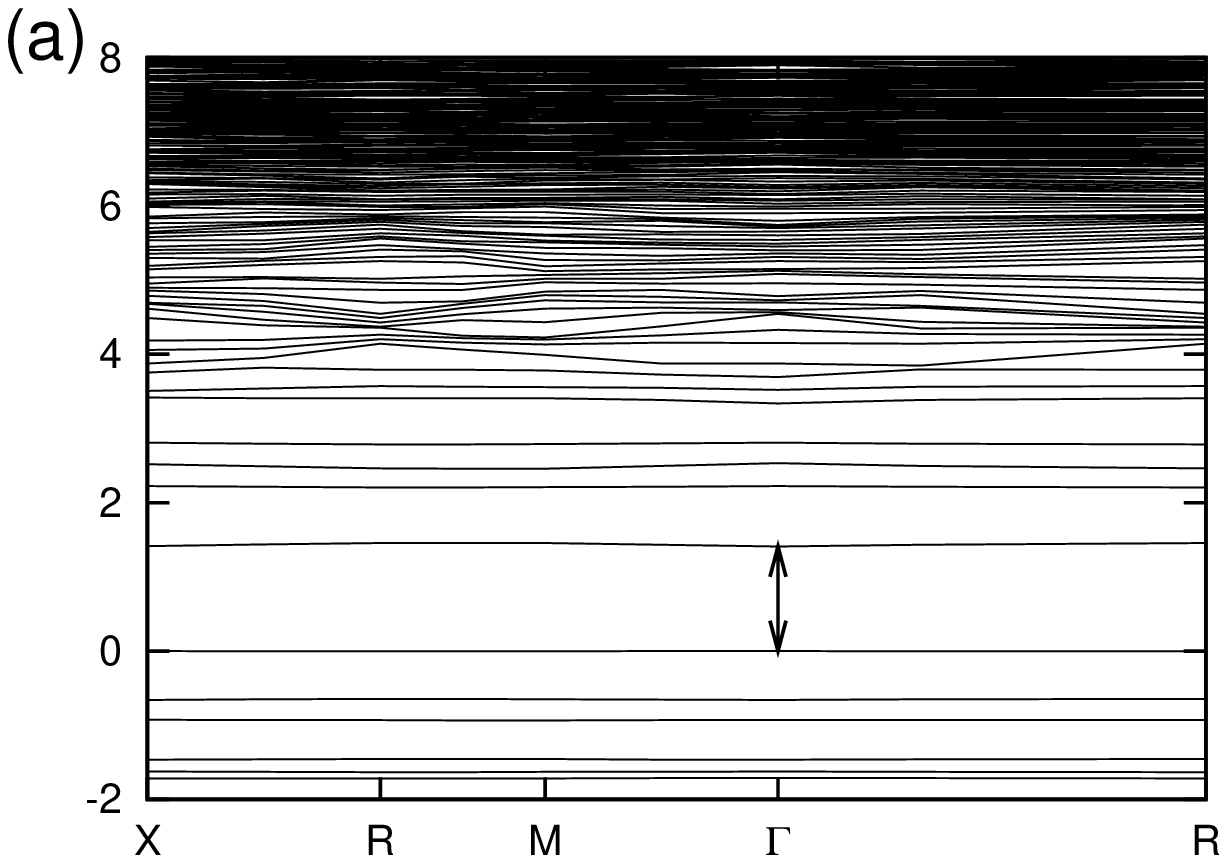,width=3cm,angle=270}
	\psfig{figure=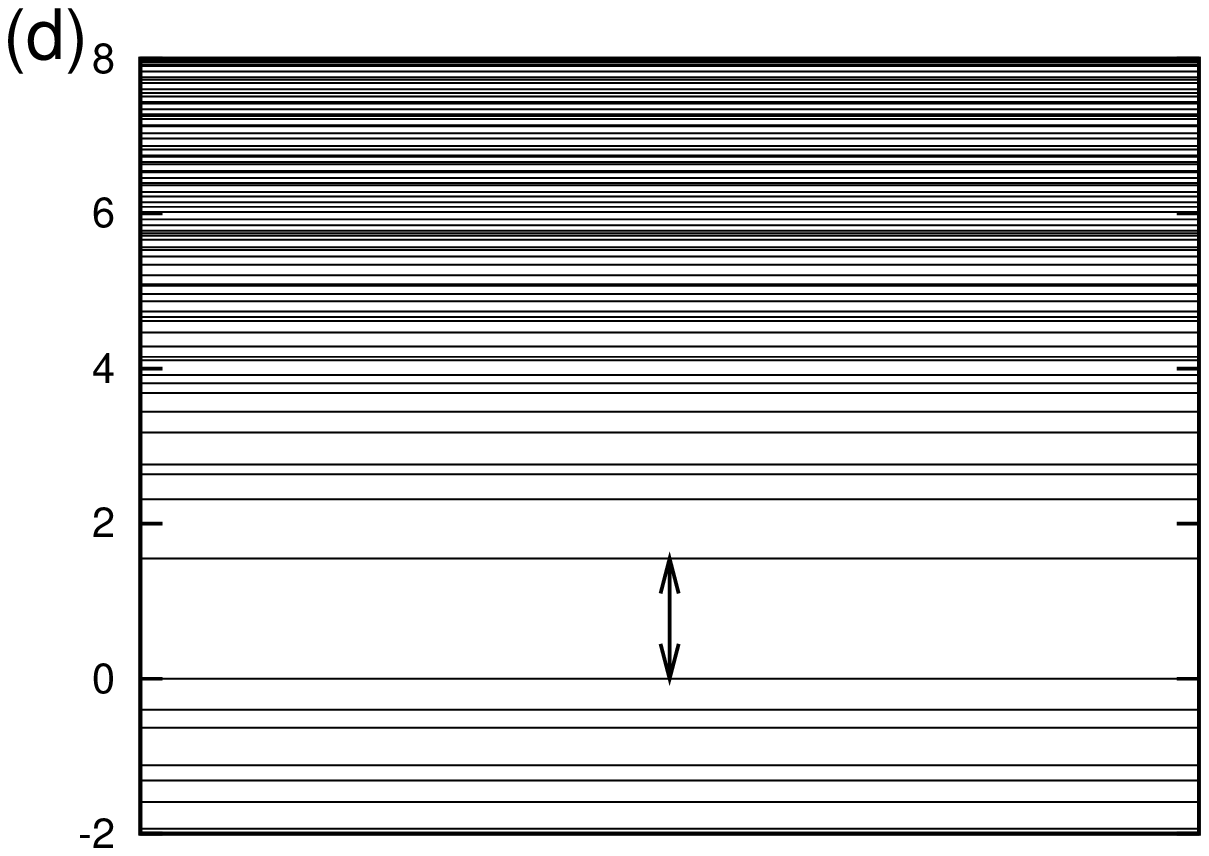,width=3cm,angle=270}}
\centerline{
	 \psfig{figure=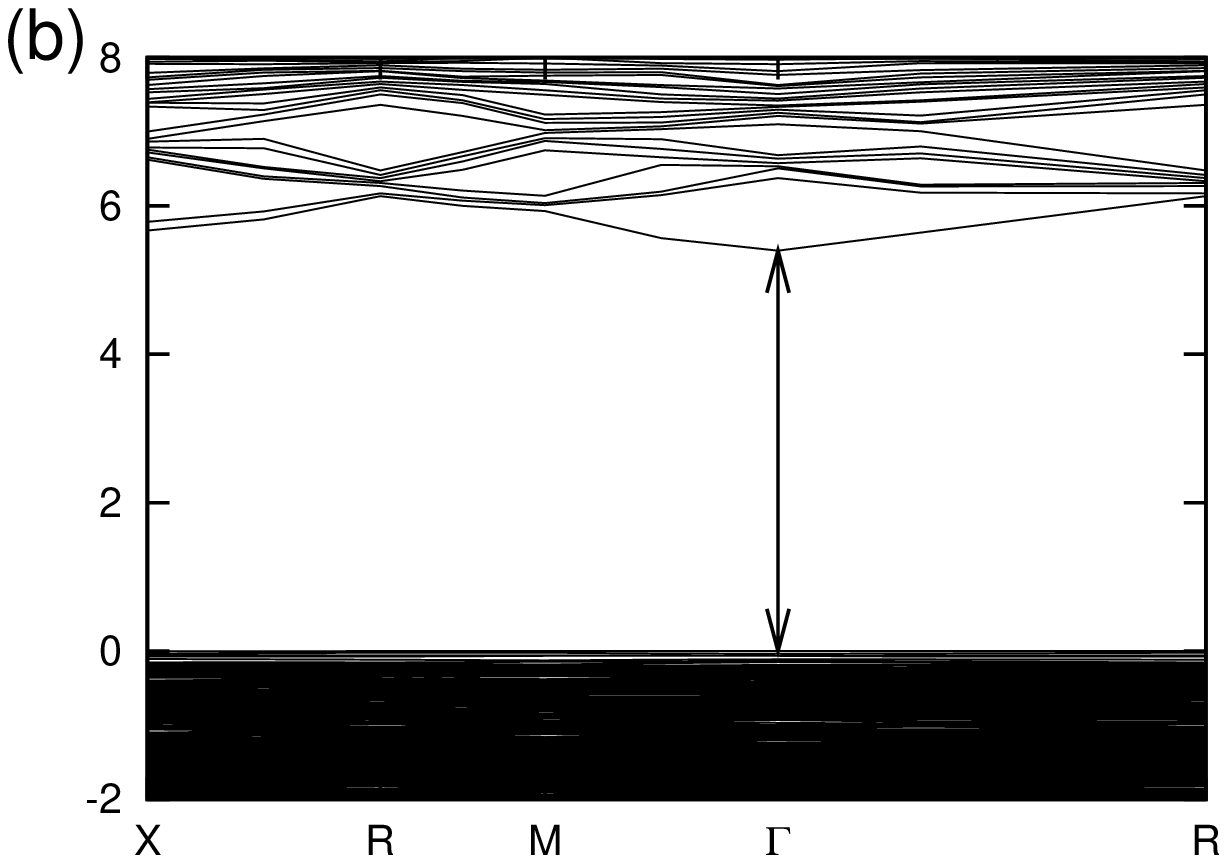,width=3cm,angle=270}
	 \psfig{figure=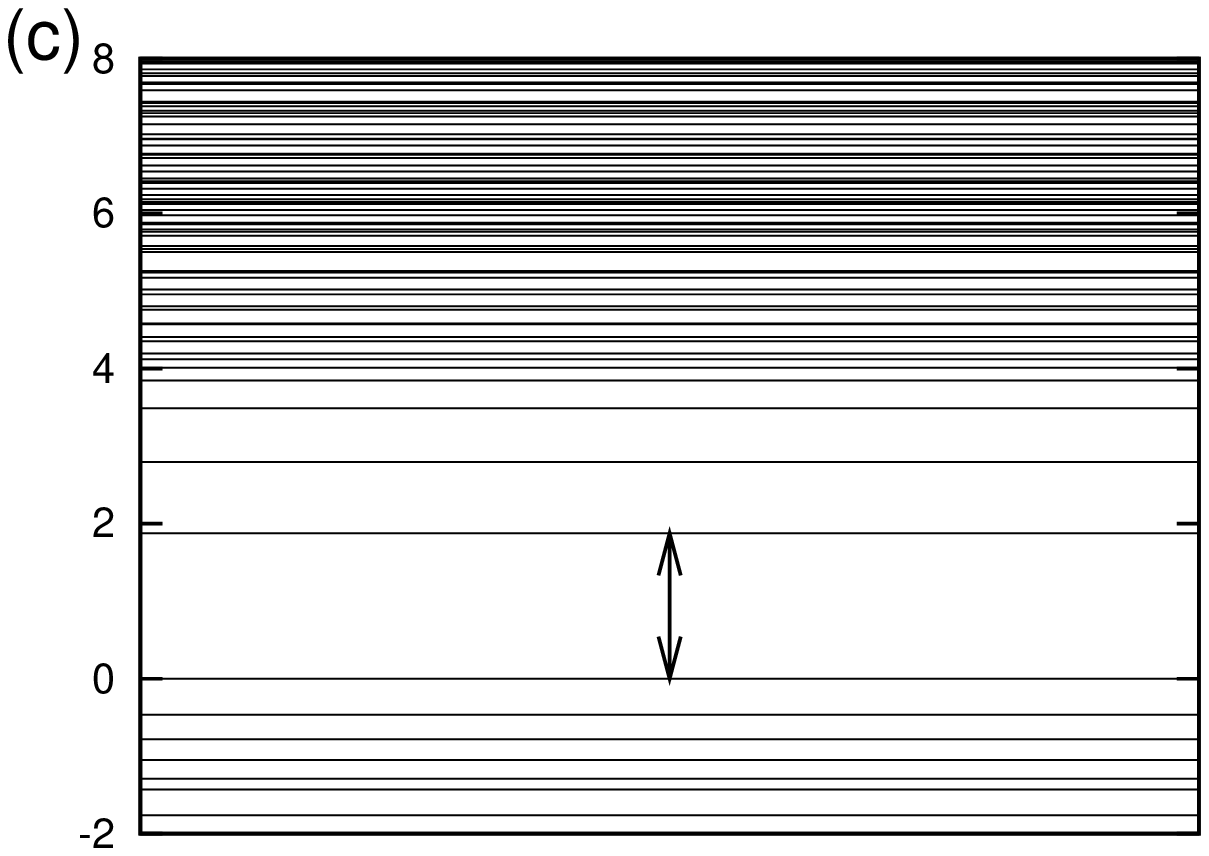,width=3cm,angle=270}}
\caption{Band structure along high symmetry points of the BZ for the a-Si$_{10}$/a-SiO$_2$ system (a)
	compared with the band structure of the glass bulk (b), the energy levels for the isolated
	a-Si$_{10}$ NC passivated by H (c) and by OH groups (d).
	The arrows show the HOMO-LUMO band gap values. The units are in eV.}
\label{vetro_bande}\end{center}\end{figure}

\begin{figure}[b]\begin{center}
\centerline{\psfig{figure=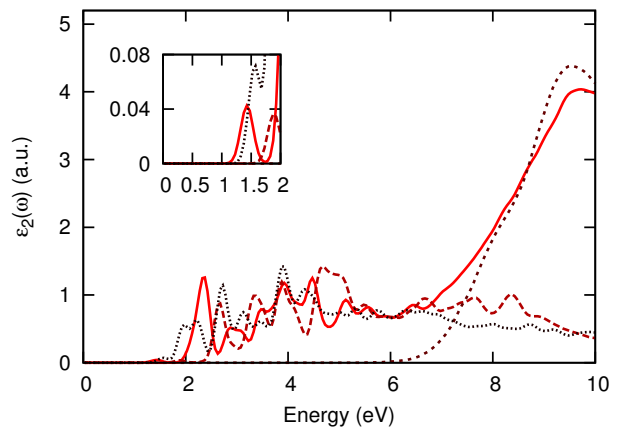,width=6cm,angle=270}}
    \caption{(color online) DFT-RPA imaginary part of the dielectric function for the a-Si$_{10}$ NC
    in a glass matrix (solid line) compared with that of isolated, hydrogenated a-Si$_{10}$-H
    NC (long-dashed line), the glass matrix (short-dashed line),
    and the hydroxided a-Si$_{10}$-OH NC (dotted line).
    The inset shows an enlargement of the spectra at low energies. }
\label{eps2_amorph}\end{center}\end{figure}

The NC, when formed in the glass, completely loses memory
of the starting tetrahedral symmetry (as shown in the previously discussed Fig. \ref{gr}).\\
No dangling bonds are present at the NC surface while some
bridge-bonded oxygens appear ({that} are not present in the
crystalline case). Despite the fact that the dramatic structural changes {with
respect to} the crystalline case, similar considerations can be
done concerning the optoelectronic properties. While the
amorphization process do{es} not affect the behavior of the glass
matrix, it strongly reduces the energy gap of the isolated
a-Si$_{10}$-H NC, of the a-Si$_{10}$-OH NC, and of the
composite a-Si$_{10}$/a-SiO$_2$ system (see Fig.
\ref{vetro_bande}).
The HOMO-LUMO gaps are summarized in table \ref{tabella_gap1}.\\
\begin{figure}[b]\begin{center}
\psfig{file=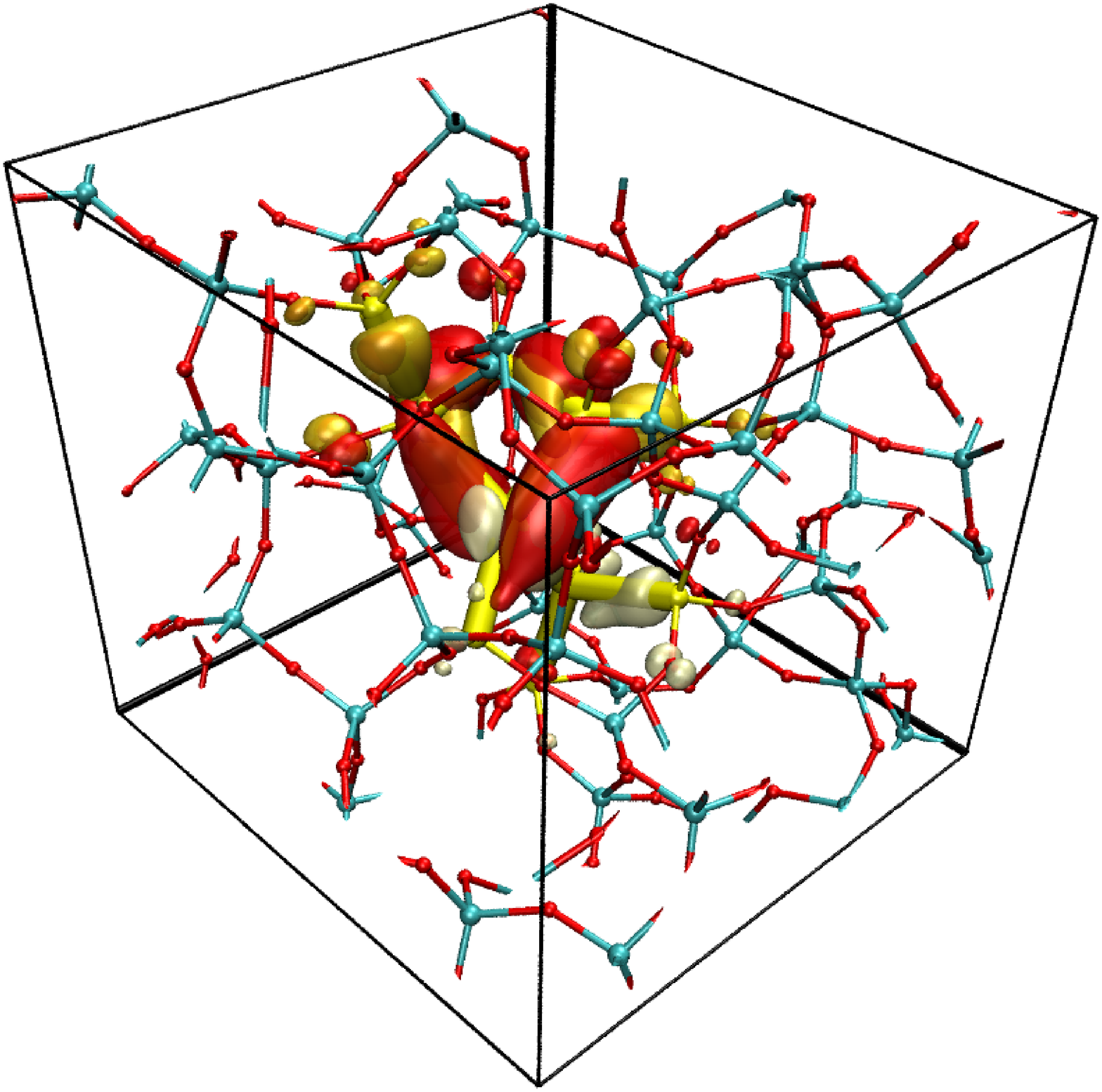,width=8cm}\\
\psfig{file=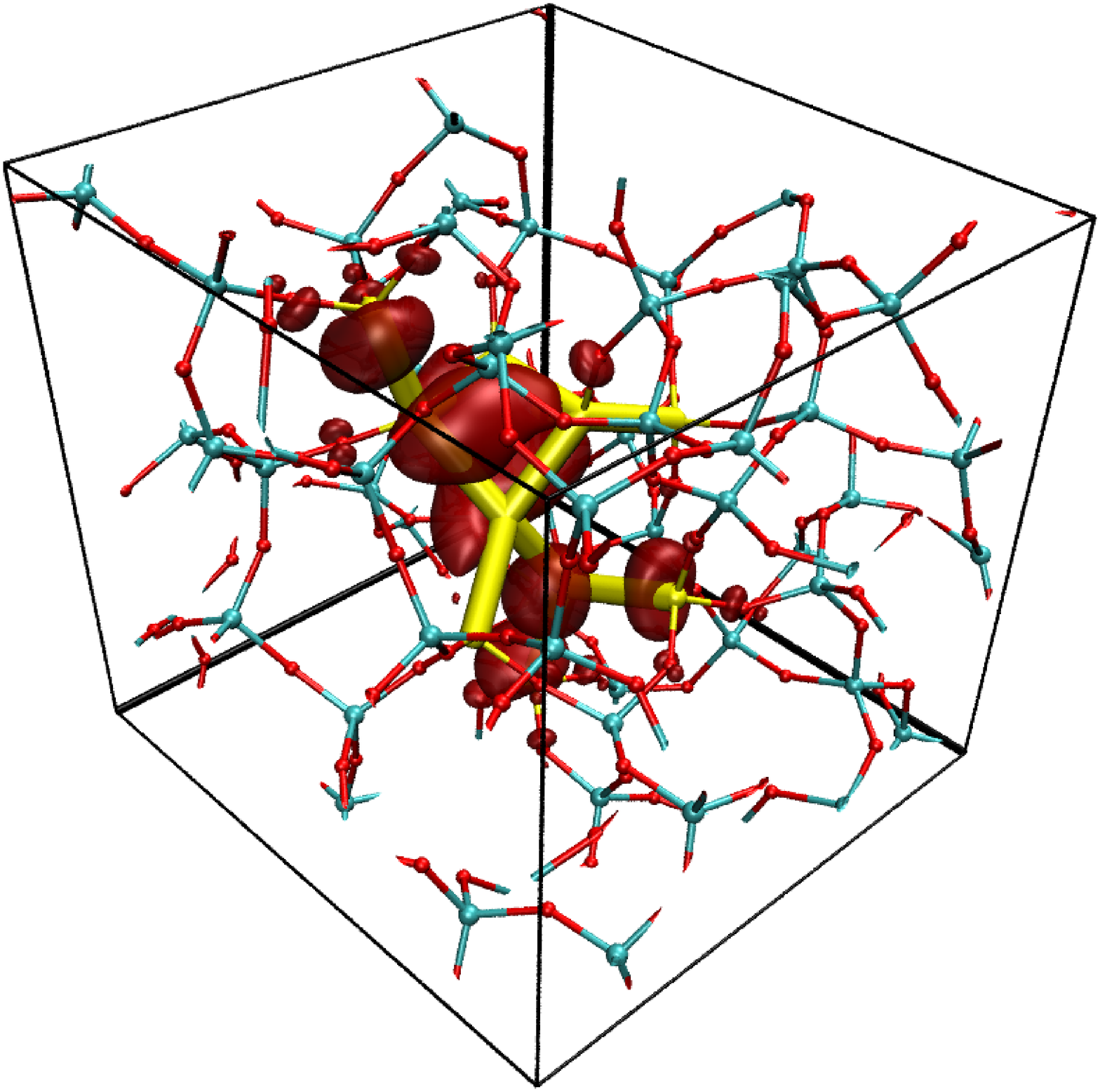,width=8cm}
\caption{(color online) Kohn-Sham orbitals at 10$\%$ of their max. amplitude
    for the a-Si$_{10}$ amorphous embedded NC.
    Top panel: HOMO state is represented in red (gray), HOMO-1 in orange (light gray),
    and HOMO-2 in white. Bottom panel: LUMO is represented in dark red (dark gray).}
\label{HLvetro}\end{center}\end{figure}
We note that, contrary to the crystalline case, the amorphous hydroxided NC
shows a gap value slightly greater than when embedded.
Also in this case the states at the valence and conduction band edges are localized at
the interface; in particular they stem from the OH passivation.
Obviously the reduction of the local symmetry induces a splitting in the energy levels
and, as a consequence, a more uniform distribution of states.\\
The HOMO, HOMO-1, HOMO-2 and LUMO states (see Fig. \ref{HLvetro})
follow the shape of the NC, resulting strongly deformed with respect to
the crystalline case, but are still clearly
confined by the interface oxygen {shell}.\\
In the calculated optical absorption spectra (see Fig. \ref{eps2_amorph})
the three regions due to the hosting matrix, Si-NC, and interface,
are still clearly distinguishable. \\
The most important differences between the calculated optical absorption spectra
for both the amorphous and the crystalline embedded NCs are {found}
in the energy range between 2 and 4 eV.
In this region the amorphous system shows more intense
peaks with respect to the crystalline case, suggesting a possible
higher emission in the visible range.\\
Despite the fact that the possibility to form amorphous NC in silica has been
recently explored\cite{flyura}, experimental measurements on
single Si-NCs with diameters of the order of 1 nm have not yet
been performed. Thus, a straightforward comparison of our results
with experimental data is not possible. Besides, the comparison
with other works\cite{Bulutay2} sustains the idea that the strong
deformation of the NC is determinant for the absorption
strength at low energies. This idea is also supported by the fact
that, for larger systems, when the shape of the NCs tends to be
spherical and the distortion is usually less
pronounced\cite{flyura2,kelires1}, crystalline and amorphous
systems produce more similar absorption spectra. This topic will
be discussed in the next subsection.

\subsection{Size Effects} \label{subsec_size}

\begin{figure}[b]\begin{center}
\centerline{\psfig{file=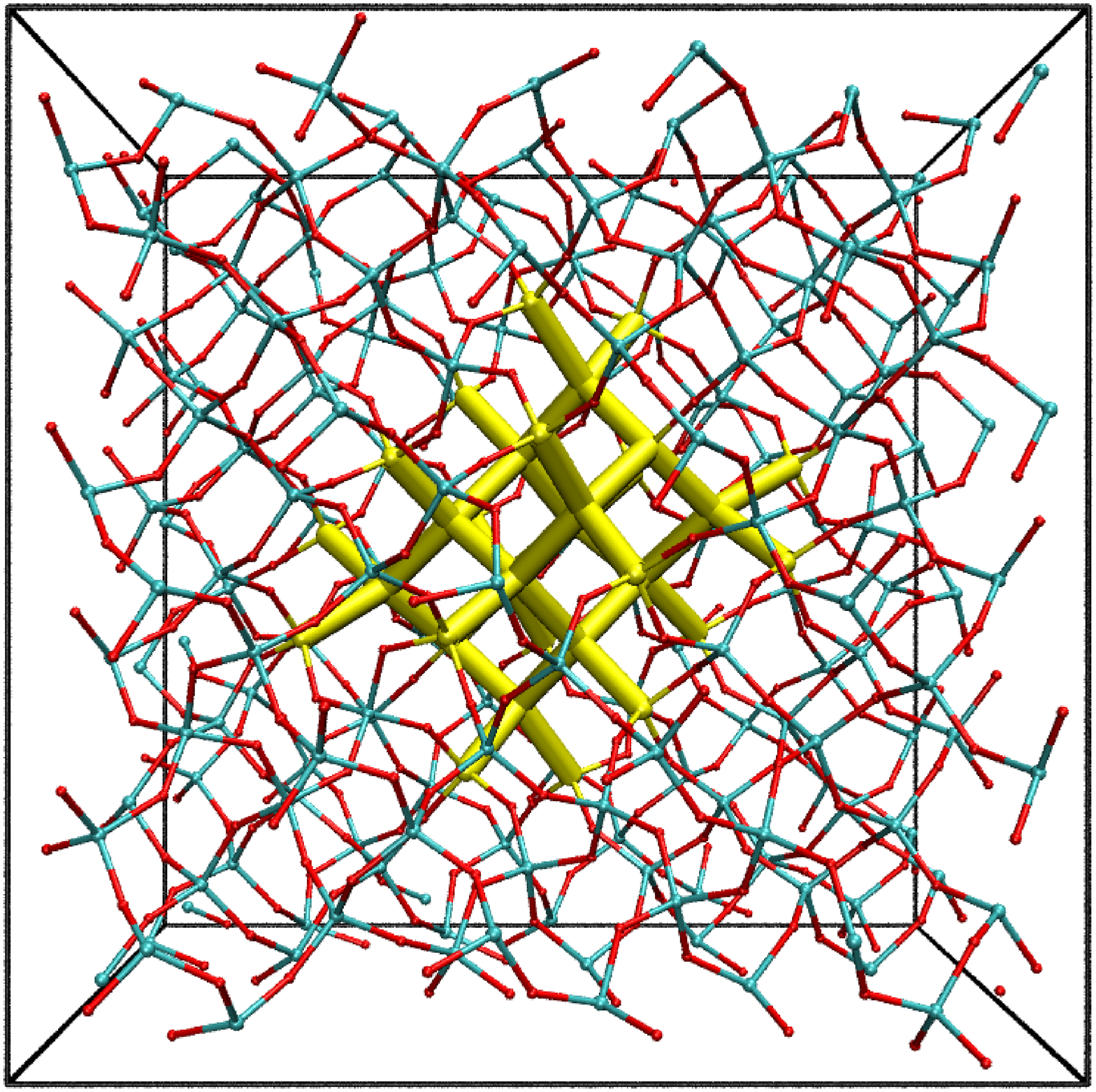,width=4.5cm}~\psfig{file=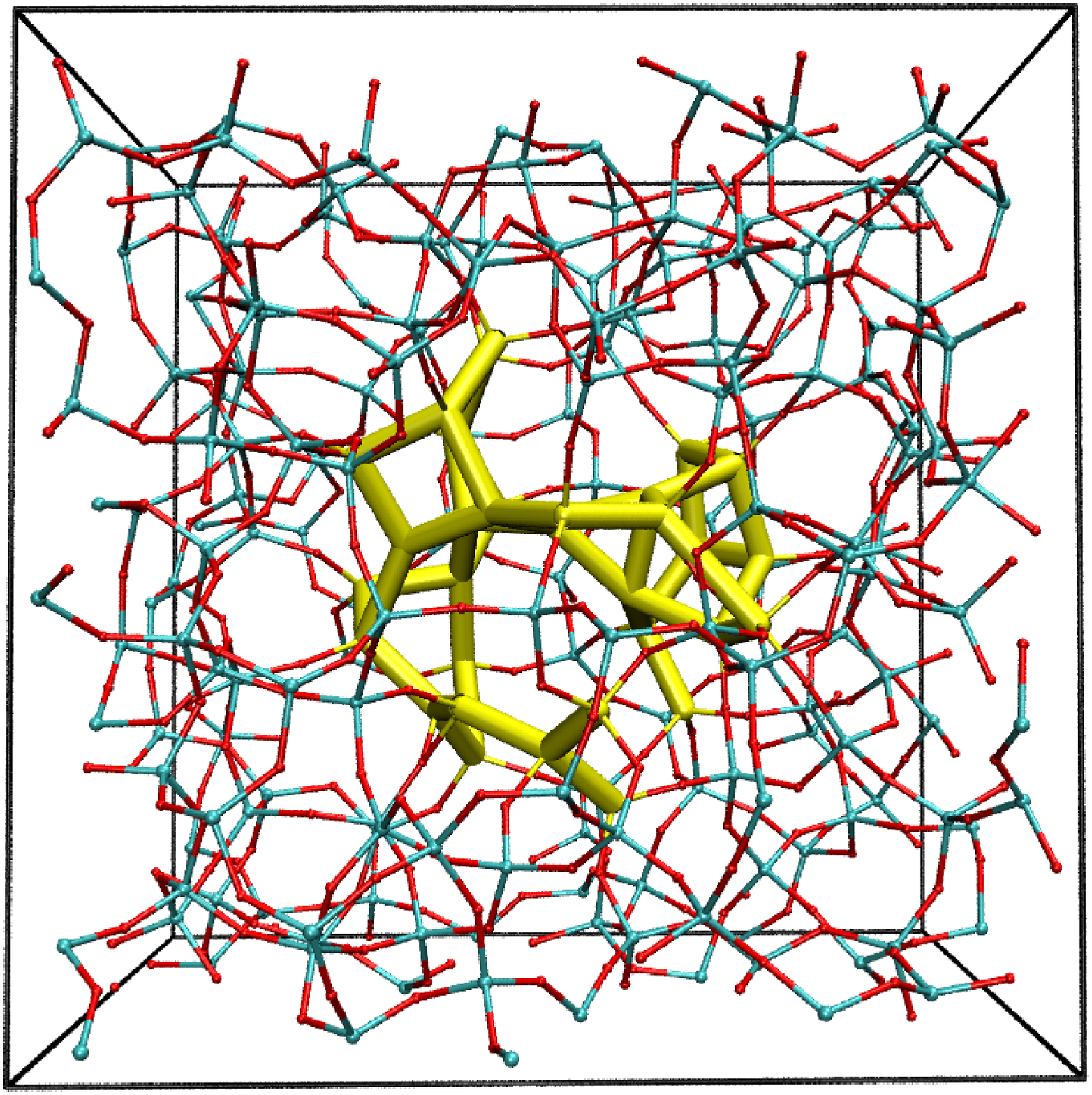,width=4.5cm}}
\caption{(color online) Stick and ball pictures of the final optimized structure of
    Si$_{32}$ NC in a $\beta$-cristobalite matrix (left) and a-Si$_{32}$ NC in a
    a-SiO$_2$ glass (right). Red (dark gray) spheres represent the O
    atoms, cyan (gray) spheres represent the Si of the matrix,
    and the yellow (gray) thick sticks represent the Si atoms of the NC.}
\label{struttura_Si32}\end{center}\end{figure}

From a theoretical point of view, the simplest model for the QC
is provided by the particle-in-a-box scheme, in which the box size is given by the NC
diameter and the potential barrier represents the host insulating matrix.
When QC effect dominates over other quantum phenomena, for an infinite barrier potential,
we have $E_G(R)=\epsilon_G+$A/R$^{\alpha}$, where $\epsilon_G$ is the bulk-silicon band gap,
R is the radius of the NC, A is a positive constant and $\alpha \leq 2$~~~\cite{lockwood}.
This model, however, is often too simple to describe correctly the experimental results.\\
Experimentally, several factors contribute to make the
interpretation of measurements of the optical gaps a difficult
task. For instance, samples show a certain distribution in the
NC size, that is difficult to be determined. In this case it is
possible that the observed PL peak does not correspond exactly to
the mean size but instead to the largest PL rate. Again, NC
synthesized by using different techniques often show different
properties in size, shape and in the interface structure.
Finally, also the NC-NC interactions can play a significant role on the emission spectrum.\\
Recent measurements on samples with diameters in the range of 2-10 nm have reported
\cite{Morante,Lieber,Kanzawa,Guha,Takeoka,Watanabe,Nassiopoulou} a PL peak that is blue-shifted
by ${\Delta}E_G(R) \sim R^{-\alpha}$, with $1 < \alpha < 1.5$.
Thus, the model described above is far from being able to reproduce the measurements.
In order to improve it, the finiteness of the potential barrier and also other corrections
should be taken into account. For example, the inclusion of the electron-hole Coulomb interaction,
that contributes a further $R^{-1}$ dependence in blue-shifting the energy, results in a more gradual
dependence of the optical gap on the dot radius, as observed experimentally.\\
As mentioned above, our investigation explores the small dot size limit, with a diameter range
within 1 nm. In this range, most of the NC atoms are positioned at the interface, where the
effects of stress and oxidation are stronger.
For bigger NCs, when Si-bulk states emerge and the surface-to-volume ratio decreases,
these effects are limited; in this case we expect a response that is independent of the
specific geometrical configuration. Thus, in order to {understand}
all the interface-related phenomena, an investigation {of} the small size limit becomes necessary.\\
To model NC of increasing size, we enlarge the hosting matrix {so
that the separation between NCs is still around 1 nm,} that we
consider enough to correctly describe the stress localized around
{each} NC and to avoid {an} overlapping {of the} NC states that
can be caused by the {application of the} periodic boundary
conditions\cite{daldosso}. These considerations lead us to
choose a BC matrix of 648 atoms (in a cubic cell with
lattice parameter of 2 nm) with a core of 32 silicon atoms,
{having} an average diameter of 1 nm. The relaxed Si$_{32}$ NC
shows a strained interface with average Si-Si bond-lengths of
$2.6$ \AA, while the silicon atoms in the core have a less
strained bond-length of $2.43$ \AA. Note that Si$_{10}$/SiO$_2$
and Si$_{32}$/SiO$_2$ have approximatively the same ratio between
the NC and the matrix volumes.
The amorphous embedded a-Si$_{32}$ NC is again produced by MD
annealing and successive {\it ab-initio} relaxation. The final
structures are shown in Fig. \ref{struttura_Si32}. Once more,
bridge-bonded oxygens are present at the interface and we find
that the number of Si-O-Si bonds increases with the dimension of
the NC (8 bonds in this case with respect to the 3 present in
the smaller NC) in nice agreement with previous results
obtained by different methods\cite{kelires1,colombo}.\\
\begin{table}[b]\begin{center} \begin{tabular}{c @{\hspace{0.6cm}} c @{\hspace{0.3cm}} c @{\hspace{0.3cm}} c}
\toprule
            & Si$_{32}$/SiO$_2$ & Si$_{32}$-OH & Si$_{32}$-H \\
\midrule
    Crystalline & 2.62              & 2.15         & 2.78 \\
    Amorphous   & 1.19              & 0.99         & 1.59 \\
\bottomrule
\end{tabular}
\caption{DFT HOMO-LUMO gap values (in eV) for the crystalline and amorphous
    silica, embedded, OH-terminated, H-terminated Si$_{32}$ NC.}
\label{tabella_gap2}\end{center}\end{table}
$E_G$ values calculated in DFT for Si$_{32}$/SiO$_2$, Si$_{32}$-OH
and Si$_{32}$-H structures are reported in Tab. \ref{tabella_gap2}.
We can note that, while $E_G$ of Si$_{32}$-H is smaller with respect
to Si$_{10}$-H due to the QC, for the other structures it strongly
depends on the interface configuration. As a consequence, for both
the  Si$_{32}$/SiO$_2$ and Si$_{32}$-OH systems, $E_G$ increases
when the phase is crystalline, and decreases when the
corresponding structures are amorphous.
We note that, while the amorphous systems present
interfaces where several types of oxidations and coordination
levels coexist, the crystalline case presents a situation in which
all the interface O atoms are single-bonded with the Si-NC atoms.
Specifically, the Si-atoms at the interface of the
Si$_{10}$/SiO$_2$ and Si$_{32}$/SiO$_2$ systems are respectively
bonded with 1.6 and 2.8 oxygens in average. It is worth to note
that a recent size-dependent experimental study of Si 2$p$
core-level shift at Si-NC/SiO$_2$ interface showed that
the shell region around the Si-NC bordered by SiO$_2$
consists of the three Si suboxide states, Si$^{1+}$, Si$^{2+}$,
and Si$^{3+}$, whose densities strongly depend on the
NC size \cite{kimkim}. Generally, such oxydation degree
is responsible for two, competitive, major effects on $E_G$: while
from one side a higher oxydation increases $E_G$, from the other
side it produces a higher strain on the Si-NC atoms that results
in a reduction of $E_G$. Briefly, what emerges is that these two
contributions are approximately of the same order of magnitude,
and the dominance of one over the other is strictly dependent on
the considered system (a publication on this topic is currently
under development).

\begin{figure}[b]\begin{center}
\centerline{\psfig{figure=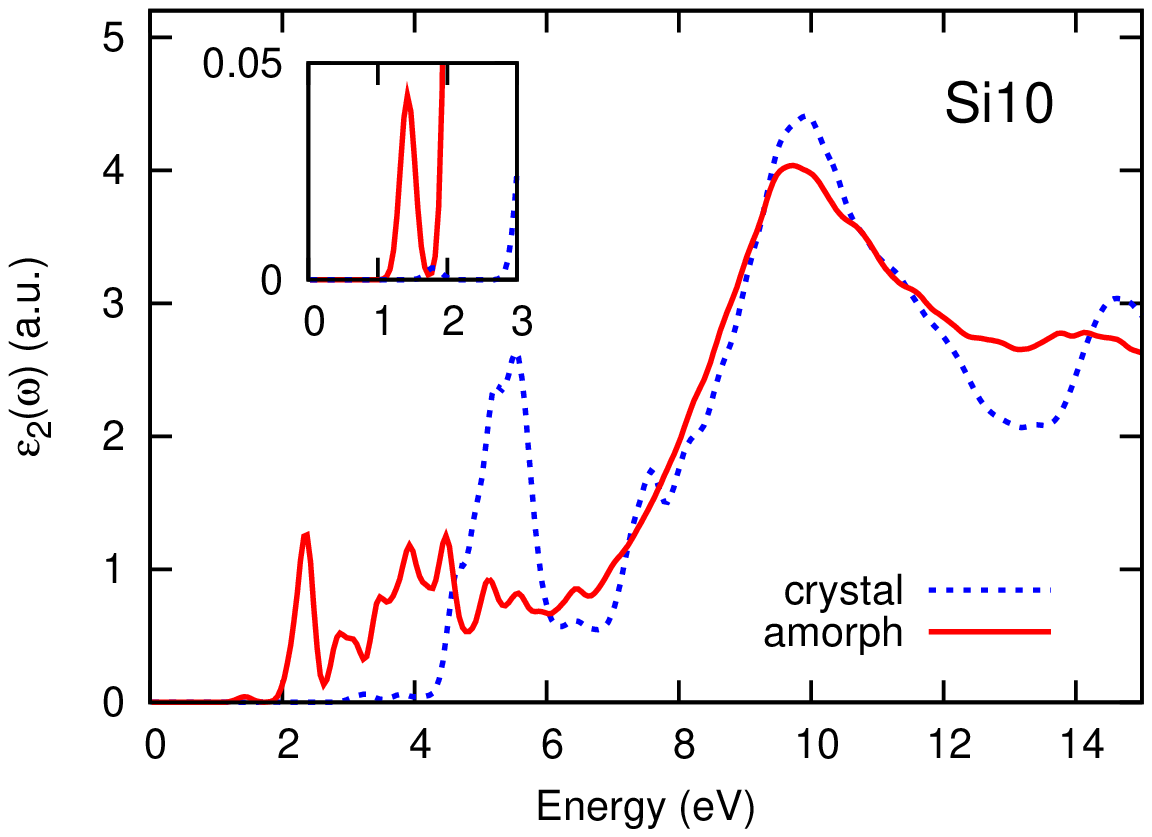,width=5cm,angle=270}}
\centerline{\psfig{figure=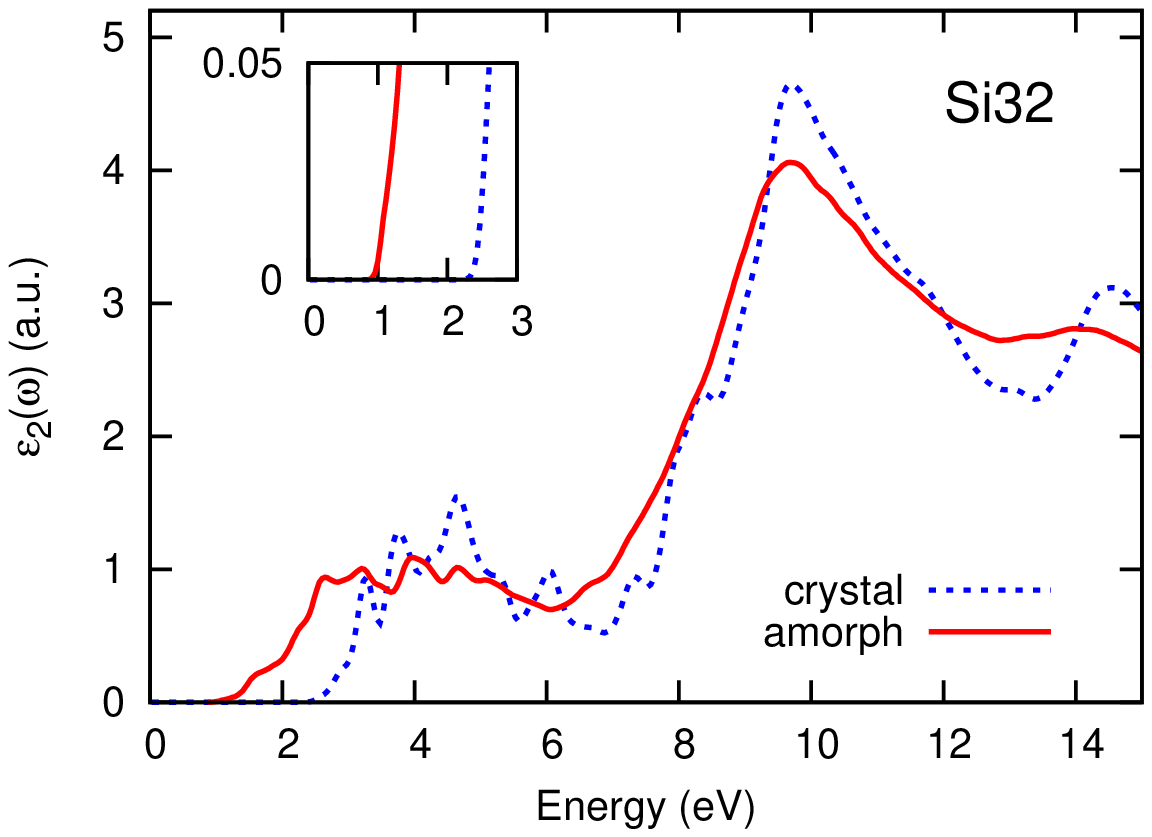,width=5cm,angle=270}}
\caption{(color online) DFT-RPA imaginary part of the dielectric function for the crystalline (dotted)
    and amorphous (solid) Si$_{10}$ (upper) and Si$_{32}$ (lower) embedded NCs.
    The insets show an enlargement of the spectra at low energies.}
\label{eps2_small-big}\end{center}\end{figure}
In our case, after performing a full relaxation, in order to remove the strain,
of the Si$_{10}$-OH and Si$_{32}$-OH structures we still obtain an $E_G$ difference of about 1 eV.
Thus, the increased $E_G$ of the larger system can be referred to the greater
amount of oxydation of the interface silicon atoms.\\
The analysis of the valence and conduction band edge states of the Si$_{32}$/SiO$_2$ system,
points out that the former are essentially localized
in a well defined part of the NC, near the interface zone, while the latter are localized on the
whole volume (from the center to the interface) of the embedded structure.\\
\begin{figure}[b]\begin{center}
\centerline{\psfig{figure=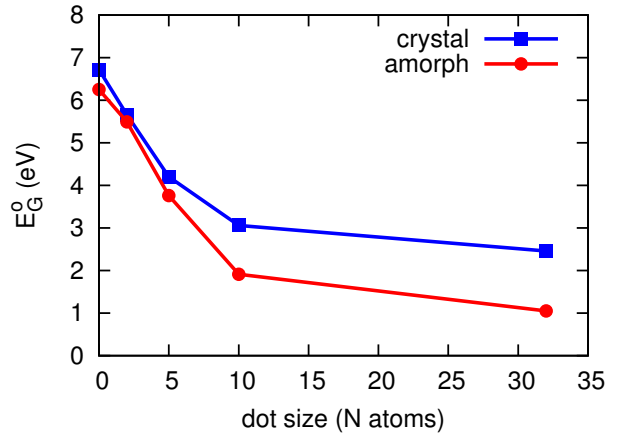,width=6cm,angle=270}}
\caption{(color online) $E_G^o$ versus the Si-NC size and amorphization (embedded systems).
    N=0 corresponds to the pure silica, N= 32, 10, 5, 2 correspond to
    an average diameters of, 1.0, 0.7, 0.45, 0.24 nm, respectively.
    The lines are drawn to guide the eye.}
\label{sizeplot}\end{center}\end{figure}
The effects induced by the amorphization process on the optical
properties {of} both Si$_{10}$/SiO$_2$ and Si$_{32}$/SiO$_2$
systems are depicted in Fig. \ref{eps2_small-big}, where the
calculated absorption spectra  for both crystalline (blue line)
and amorphous (red line) phases are reported. Here we note that,
while in the smaller Si$_{10}$/SiO$_2$ system the amorphization
{produces} a net change in the shape of the spectrum below 6 eV,
inducing also new strong structures in the energy windows between
2 and 4 eV, for the bigger Si$_{32}$/SiO$_2$  it seems to induce
{only a} simple red-shift. These facts enforce our previous
hypothesis that for bigger NCs the particular configuration of the
interface becomes less important. This is especially true for
{the} amorphous systems, where the diversity of {the} atomic
configurations tend to average {out} the final properties. In
order to compare theoretical and experimental results for the
energy gap dependency, we introduce a new quantity, $E_G^o$,
which is defined as the optical absorption threshold when
all the transitions with an intensity lower than 1\% of the
highest peak are neglected. In this way we introduce a sort of
"instrument resolution" by neglecting very low oscillator
strength optical transitions (for instance the HOMO-LUMO
transition in the Si$_{10}$/SiO$_2$ crystalline system).
The obtained DFT $E_G^o$ values for the embedded systems are
depicted in Fig. \ref{sizeplot}. Here $E_G^o$ has been
calculated for Si-NC containing 2, 5, 10 and 32 atoms. It is
clear that, while the increasing of the NC size always results
in a reduction of $E_G^o$, the effect of the amorphization is
to introduce an additional red-shift that becomes relevant for large NCs.
For large number of atoms N we expect however a vanishing of the
red-shift due to the fact that, for semiconductors, the main
optical properties should only depend on the short range-order\cite{Tauc}.

\section{Excitonic and Many-Body Corrections}\label{sec_EXC}
\begin{figure}[t]\begin{center}
\centerline{\psfig{file=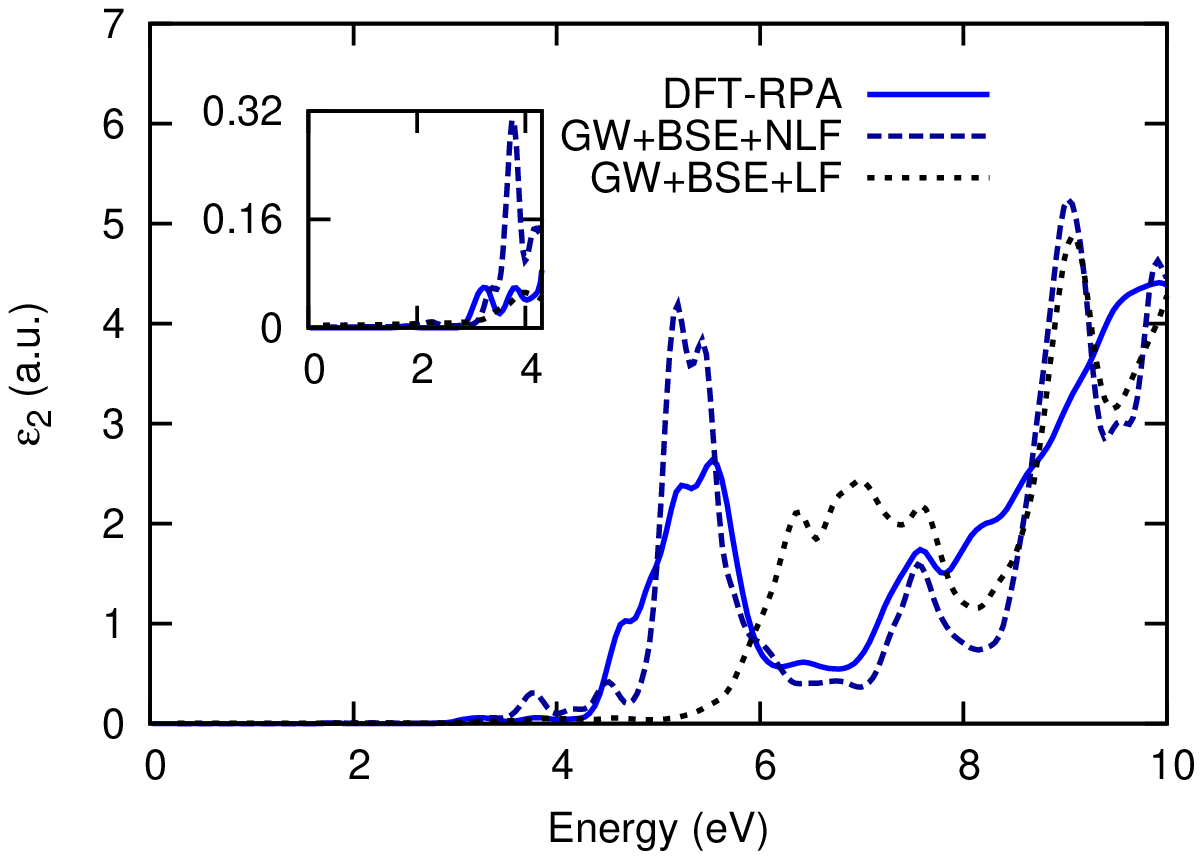,width=5.2cm,angle=270}}
\centerline{\psfig{file=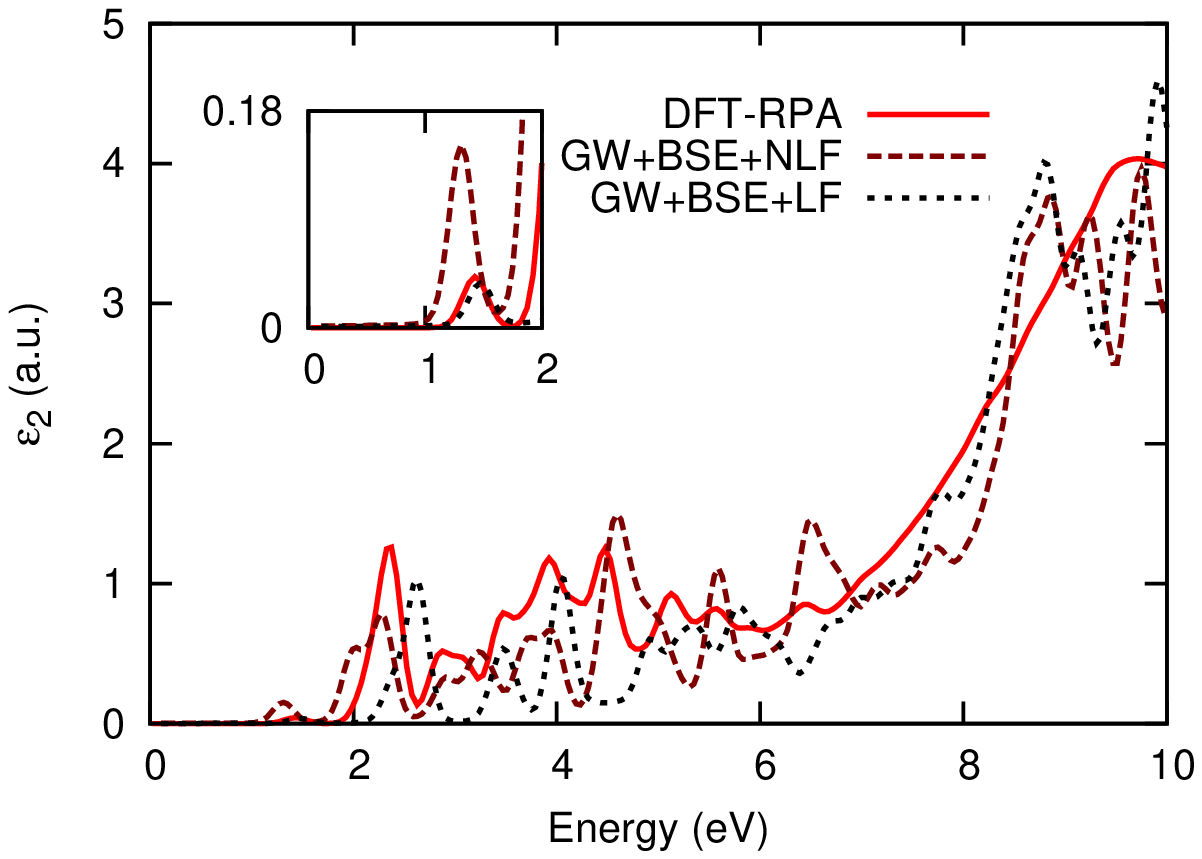,width=5.2cm,angle=270}}
\caption{(color online) DFT-RPA, GW+BSE+NLF (that is, without local fields),
    and GW+BSE+LF (with local fields) calculated imaginary part of the
    dielectric function for crystalline (top) and amorphous (bottom) case.}
\label{eps2_gwbse}\end{center}\end{figure}

Self-Energy  and excitonic effects are known to play a very
important role both in low dimensional systems (as the quantum
dots) and in 3-dimensional systems as {the} SiO$_2$. In this last
case it is known that DFT underestimates the electronic gap of
about 5 eV, and important  excitonic effects are responsible of the
strong absorption peak at about 10 eV. The inclusion of the
many-body effects is thus of fundamental importance in order to
obtain a better description of the optical properties of NCs
embedded in SiO$_2$. Quasi-particle effects within the GW approach
\cite{Hedin} and excitonic effects within the BS equation
\cite{reviewLucia} have been considered in the calculation of
the Si$_{10}$/SiO$_2$ electronic properties.\\
Table \ref{tabella_gap3} shows the calculated $E_G$ within DFT,
GW, GW+BSE+LF approximations. It is possible to observe that,
in the ordered (amorphous) case, the inclusion of the GW corrections
opens up the gap by about 1.9 (1.7) eV, while the excitonic and
LF correction reduces it by about 1.5 (1.6) eV. Thus, the total
correction to the LDA $E_G$ results to be of the order of 0.4 (0.1) eV.
We note that the final $E_G$ of the crystalline and amorphous embedded
NCs are quite different, i.e. 2.17 and 1.51 eV, respectively.\\
\begin{table}[b]\begin{center}\begin{tabular}{c @{\hspace{0.6cm}} c @{\hspace{1cm}} c @{\hspace{0.5cm}} c}
\toprule
		    & DFT   & GW    & GW+BSE+LF \\
\midrule
	Crystalline & 1.77  & 3.67  & 2.17 \\
	Amorphous   & 1.41  & 3.11  & 1.51 \\
\bottomrule
\end{tabular}
\caption{Many-body effects on the gap values (in eV) for the
    crystalline and amorphous embedded Si$_{10}$ dots.}
\label{tabella_gap3}\end{center}\end{table}
The difference between the GW electronic gap and the GW+BSE+LF
optical excitonic gap gives the exciton binding energy E$_b$.
Our calculated exciton binding energies are quite large: 1.5 eV
(crystalline) and 1.6 eV (amorphous). They are very large if
compared with that of bulk SiO$_2$ (almost 0 eV)~\cite{chang0,marini,Bechstedt_betacristo},
bulk Si ($\sim$ 15 meV) or with carbon nanotubes \cite{spataru,chang}
where E$_b\sim$\,1\,eV, but {they are} similar
to those calculated for undoped and doped Si-NC \cite{leo,feffe0}
of similar size and for Si and Ge small nanowires \cite{bruno1,bruno2}.\\
Fig. \ref{eps2_gwbse} shows the calculated DFT-RPA, GW+BSE+NLF, and GW+BSE+LF
absorption spectra for the embedded Si$_{10}$ and a-Si$_{10}$ NCs.
The results show that the inclusion of the many-body effects does not
substantially modify the absorption spectra. In both cases the energy
position of the absorption onset is practically not modified (see insets).
Delerue and coworkers\cite{Delerue} found that, for Si-NC larger
than 1.2 nm, the self-energy and Coulomb corrections almost cancel
each other. In our case (about 0.7nm), this cancellation
does not completely occur, especially for the crystalline system, where
self-energy effects dominate.\\
Finally, we note that the inclusion of LF contributions has dramatic effects
on the spectrum of the crystalline system, especially in the region between 4 and 6 eV,
while the amorphous system seems to be affected by a simple blue-shift of the absorption threshold.

\section{Conclusions} \label{sec_concl}
Considering as the $\beta$-cristobalite phase of the SiO$_2$, like as the
amorphous phase obtained by classical molecular-dynamics, we have generated
silicon nanoclusters of different size and shape, that are embedded in a silica matrix.
The final structures, that have been relaxed with the SIESTA method\cite{siesta1},
shows that the surrounding matrix always produces some strain on the nanocluster,
especially at the Si/SiO$_2$ interface.
Then we have compared the electronic and optical properties of the embedded structures,
calculated within the DFT-LDA, with that of the free H- and OH- terminated clusters,
and that of the pure silica.
What emerges is that the electronic and optical behavior is strongly influenced
by the structural properties of the considered systems; actually, apart for the
quantum confinement effect on the energy gap due to the different
NCs size, the amorphization of the embedded NC is responsible for
the main electronic and optical changes of the whole system.\\
For the smaller cases, we have also investigated the local field, self energy, and binding energy effects.
Concerning the optical properties we have shown that, while the many
body approach provides only minor corrections, the inclusion of local
fields is crucial in the analysis of highly symmetric systems
(crystalline case) but it is of minor importance in disordered structures (amorphous case).\\
This work shows that it could be worthy to experimentally investigate
the optical properties of the matrix embedded amorphous nanoclusters\cite{pavesi2,Khriachtchev},
that show different properties with respect to their crystalline counterparts.\\
This work is supported by PRIN2007 and CNR Italia-Turchia. We
acknowledge CINECA CPU time granted by INFM (Progetto Calcolo
Parallelo). We thanks A. Pedone for the contribution with the
modelling of the silica glasses.

\end{document}